\documentclass[useAMS,usenatbib]{mn2e}
\usepackage{graphicx}
\usepackage{amsmath}
\usepackage{amssymb}

\newcommand\nodata{ ~$\cdots$~ }%

\newcommand\aj{{AJ}}%
\newcommand\araa{{ARA\&A}}%
\newcommand\apj{{ApJ}}%
\newcommand\apjl{{ApJ}}%
\newcommand\apjs{{ApJS}}%
\newcommand\aap{{A\&A}}%
\newcommand\aaps{{A\&AS}}%
\newcommand\mnras{{MNRAS}}%
\newcommand\pasp{{PASP}}%
\newcommand\pasj{{PASJ}}%
\title[Supernovae in the SDF: An Initial Sample, and Type Ia Rate, Out to $z\approx1.6$]
{Supernovae in the Subaru Deep Field:\\
 An Initial Sample, and Type Ia Rate, out to Redshift 1.6}
\author[Poznanski et al.]
{D. Poznanski$^{1}$\thanks{E-mail: dovip@wise.tau.ac.il},
D. Maoz$^{1,2}$,
N. Yasuda$^{3}$,
R. J. Foley$^{4}$,
M. Doi$^{5}$,
A. V. Filippenko$^{4}$,\newauthor
M. Fukugita$^{3}$,
A. Gal-Yam$^{6,7,2}$,
B. T. Jannuzi$^{7}$,
T. Morokuma$^{9}$,
T. Oda$^{9}$,
H. Schweiker$^{10}$,\newauthor
K. Sharon$^{1}$,
J. M. Silverman$^{4}$,
and T. Totani$^{11}$\\
\\
$^{1}$School of Physics and Astronomy, Tel-Aviv University, Tel-Aviv 69978, Israel.\\
$^{2}$Kavli Institute for Theoretical Physics, University of California, Santa Barbara, CA 93106-4030, USA.\\
$^{3}$Institute for Cosmic Ray Research, University of Tokyo, Kashiwa 277-8582, Japan.\\
$^{4}$Department of Astronomy, University of California, Berkeley, CA 94720-3411, USA.\\
$^{5}$Institute of Astronomy, University of Tokyo, Mitaka, Tokyo 181-8588, Japan.\\
$^{6}$Astrophysics Group, Physics Faculty, Weizmann Institute of Science,Rehovot 76100, Israel.\\
$^{7}$Astronomy Department, California Institute of Technology, Pasadena, CA 91125, USA.\\
$^{8}$National Optical Astronomy Observatory, Tucson, AZ 85726-6732, USA.\\
$^{9}$National Astronomical Observatory of Japan, 2-21-1 Osawa, Mitaka, Tokyo 181-8588, Japan.\\
$^{10}$WIYN Consortium, Inc., 950 North Cherry Avenue, Tucson, AZ 85719, USA.\\
$^{11}$Department of Astronomy, School of Science, Kyoto University, Sakyo-ku, Kyoto 606-8502, Japan.\\}
\begin{document}
\maketitle
\label{firstpage}
\begin{abstract}

Large samples of high-redshift supernovae (SNe) are
potentially powerful probes of cosmic star formation, metal
enrichment, and SN physics.
We present initial results from a new deep SN survey,
based on re-imaging in the $R,~i',~z'$ bands, of the $0.25~\mathrm{deg}^2$
Subaru Deep Field (SDF), with the 8.2~m Subaru telescope and Suprime-Cam.
In a single new epoch consisting of two nights of observations, we have
discovered 33 candidate SNe, down to a $z'$-band magnitude of 26.3 (AB).
We have measured the
photometric redshifts of the SN host galaxies, obtained Keck
spectroscopic redshifts for 17 of the host galaxies, and classified
the SNe using the Bayesian photometric algorithm of Poznanski et
al. (2007) that relies on template matching.
After correcting for biases in the classification, 55\% of our
sample consists of Type Ia supernovae and 45\% of core-collapse SNe. The redshift
distribution of the SNe~Ia reaches $z \approx 1.6$, with a median
of $z \approx 1.2$. The core-collapse SNe reach $z \approx 1.0$, with
a median of $z \approx 0.5$.  Our SN sample is comparable to the
\textit{Hubble Space Telescope}/GOODS sample both in size and redshift
range.
The redshift distributions of the SNe in the SDF and in GOODS are consistent,
but there is a trend (which requires confirmation using a larger sample)
for more high-$z$ SNe~Ia in the SDF. This trend is also apparent when comparing
the SN~Ia rates we derive to those based on GOODS data. Our results suggest a
fairly constant rate at high redshift that could be tracking the star-formation rate.
Additional epochs on this field, already being obtained, will enlarge our SN sample
to the hundreds, and determine whether or not there is a decline in the
SN~Ia rate at $z \gtrsim 1$.

\end{abstract}


\section{Introduction}

Supernovae (SNe) hold the key to several open astrophysical questions.
In particular, SNe are the main source of heavy elements,
and their energy input to the interstellar medium (ISM) is a
vital ingredient in galaxy formation \nocite{Yepes_97}(e.g., {Yepes} {et~al.} 1997).
Learning the nature of the progenitors of different SN types, the SN
rates, and the SN distributions in both space and cosmic time, are
essential steps toward understanding metal enrichment and galaxy
formation \nocite{Kobayashi_00}(e.g.,  {Kobayashi}, {Tsujimoto} \&  {Nomoto} 2000), and will be a prime topic of
study with the James Webb Space Telescope.

It is widely believed that SNe~Ia arise from the explosions of white
dwarfs (WDs) in binary systems, while all other SN types result from
massive-star core collapse \nocite{FILIPPENKO_97, AGY_05gl,li_07}(e.g., {Filippenko} 1997; {Gal-Yam} {et~al.} 2007; {Li} {et~al.} 2007, and references
therein).  Core-collapse SNe (CC-SNe) therefore
explode promptly ($\lesssim 10$~Myr) after the formation of a stellar
population, and the CC-SN rate will trace the star-formation history
(SFH).
In contrast, SNe~Ia should occur only after WD formation and
binary evolution, with a ``delay'' of order 0.1--10~Gyr. The SN~Ia
rate vs. cosmic time will therefore be a convolution of the SFH with a
``delay function'' --- the SN~Ia rate following a brief burst of star
formation.

The progenitors of SNe~Ia have not been identified, and could be WDs
accreting from main-sequence companions, sub-giants, or giants, or
alternatively ``double-degenerate'' WD pairs that merge
\nocite{Hilleb_IaSNe}(e.g., {Hillebrandt} \& {Niemeyer} 2000, and references therein).
Each of these scenarios predicts a different delay function (e.g.,
\nocite{Kobayashi_98}{Kobayashi} {et~al.} 1998; \nocite{Madau_98}{Madau}, {Della Valle} \&  {Panagia} 1998; \nocite{Yungelson_98}{Yungelson} \& {Livio} 1998;
\nocite{belczynski_05}{Belczynski}, {Bulik} \&  {Ruiter} 2005; \nocite{greggio_05}{Greggio} 2005; see recent updates in \nocite{Panagia_07}{Panagia}, {Della Valle} \&  {Mannucci} 2007),
which in turn dictates the SN rate vs. redshift. For the local
Universe, current estimates of SN rates \nocite{Cappellaro_99,mannucci_05,Sharon_07}(e.g.,
 {Cappellaro}, {Evans} \&  {Turatto} 1999; {Mannucci} {et~al.} 2005; {Sharon} {et~al.} 2007) are still uncertain, but
improving substantially (Leaman, Li, \& Filippenko 2007, in preparation). At
$z \approx 0.3-1.6$, a number of sometimes conflicting measurements
exist \nocite{Pain_02,AGY_02,TONRY_03,
DAHLEN_SNR04,Cappellaro_05,BARRIS_SNR06,Neill_06,Sullivan_06}({Pain} {et~al.} 2002; {Gal-Yam}, {Maoz} \& {Sharon} 2002; {Tonry} {et~al.} 2003; {Dahlen} {et~al.} 2004; {Cappellaro} {et~al.} 2005; {Barris} \& {Tonry} 2006; {Neill} {et~al.} 2006; {Sullivan} {et~al.} 2006b).

Recent attempts to constrain the cosmic SFH and the SN~Ia delay
function using SN rates have yielded some intriguing
results. \nocite{AGY_02}{Gal-Yam} {et~al.} (2002) carried out the first measurement of cluster
SN~Ia rates at $z = 0.25$ and $z = 0.9$ using deep ($R \approx
27$~mag) archival \textit{Hubble Space Telescope (HST)} images of
eight massive clusters. In \nocite{MAOZ_AGY_04}{Maoz} \& {Gal-Yam} (2004), the measured rates
were compared to predictions normalized by the requirement that SNe~Ia
produce the observed mass of iron in clusters.
The low observed rate at $z \approx 1$ implies that the SNe needed
to produce the iron exploded at earlier times, arguing against
models with long SN~Ia time delays, which predict SN rates ten
times higher than observed.
Thus, cluster
iron production by SNe~Ia appeared to be a viable option only if
SNe~Ia have relatively short ($\lesssim 2$~Gyr) time delays. In contrast,
comparison of the SN~Ia rate vs. $z$ in the {\it field} to the cosmic
SFH has generally suggested a {\it long} delay time
(\nocite{Pain_02}{Pain} {et~al.} 2002; \nocite{TONRY_03}{Tonry} {et~al.} 2003; \nocite{AGY_MAOZ_04}{Gal-Yam} \& {Maoz} 2004;
\nocite{STROLGER_04}{Strolger} {et~al.} 2004; see, however, \nocite{BARRIS_SNR06}{Barris} \& {Tonry} 2006, who find
evidence for a short delay). These results may therefore point to
CC-SNe from an early generation of star formation with a top-heavy
initial mass function, rather than SNe~Ia, as the source of cluster
enrichment.

However, a re-analysis by \nocite{Forster_06}{F{\"o}rster} {et~al.} (2006) of the {\it HST}-based
Great Observatories Origins Deep Survey (GOODS) data of
\nocite{STROLGER_04}{Strolger} {et~al.} (2004), originally obtained to measure the expected early
deceleration of the Universe \nocite{RIESS_04}({Riess} {et~al.} 2004a), indicates that these
conclusions may be strongly dependent on the assumed SFH.
Furthermore, evidence has surfaced that there may be two separate
SN~Ia channels, a ``prompt'' ($\sim 1$~Gyr) channel and a ``tardy''
($\sim 10$~Gyr) one \nocite{mannucci_05,Scannapieco_05,Mannucci_06}({Mannucci} {et~al.} 2005; {Scannapieco} \& {Bildsten} 2005; {Mannucci}, {Della Valle} \&  {Panagia} 2006).  A
single delay distribution that is peaked at short delays but includes
a long tail extending to large delays may be able to reproduce most of
the current data. Nevertheless, some rate measurements at similar
redshifts are discrepant by large factors (e.g., \nocite{Pain_02}{Pain} {et~al.} 2002
vs. \nocite{BARRIS_SNR06}{Barris} \& {Tonry} 2006). There is ongoing debate as to whether
incompleteness or contamination of SN samples are affecting the low or
high-rate measurements, respectively. The decrease in the field SN~Ia rate
at $z>1$ measured by \nocite{DAHLEN_SNR04}{Dahlen} {et~al.} (2004) and \nocite{STROLGER_04}{Strolger} {et~al.} (2004), if
real, does not permit the existence of the prompt SN~Ia component
indicated by the high SN~Ia rate seen in star-forming galaxies
\nocite{mannucci_05,Mannucci_06,Sullivan_06}({Mannucci} {et~al.} 2005, 2006; {Sullivan} {et~al.} 2006b). \nocite{Scannapieco_05}{Scannapieco} \& {Bildsten} (2005) have actually
predicted that this apparent decrease will disappear with improved
data.  \nocite{BARRIS_SNR06}{Barris} \& {Tonry} (2006) have argued that the claims for long delay times depend strongly 
on the presence of very few SNe at the highest redshifts in the
GOODS survey. Indeed, small-number statistics are a major limiting
factor in all current estimates of SN rates at high $z$.  Progress in
this field requires larger high-$z$ SN samples than presently
available.

However, with typical magnitudes of $\sim$25--26, samples of high-$z$ SNe
are currently difficult to confirm (or classify) spectroscopically.
Photometric classification of SNe has been shown do be a viable option
by \nocite{POZ_TP1}{Poznanski} {et~al.} (2002), and further explored since then \nocite{AGY_TP2,RIESS_TP,SULL_TP}(e.g.,  {Gal-Yam} {et~al.} 2004b; {Riess} {et~al.} 2004b; {Sullivan} {et~al.} 2006a).
Recently, \nocite{TP3}{Poznanski}, {Maoz} \& {Gal-Yam} (2007, hereafter P07) have developed an automatic Bayesian
algorithm for the purpose of classifying SNe with single-epoch photometry.
The major novelties in the approach of P07 lie in the probabilistic
classification, and the ability to quantify the success fractions, 
uncertainties, and biases of photometric SN typing.
This opens the possibility for surveys for large numbers of SNe,
and their classification, using solely single-epoch, ground-based, imaging.

In this paper, we present initial results from a survey in which we
attempt this new approach for obtaining large high-$z$ SN samples. The
survey is conducted by re-imaging the Subaru Deep Field \nocite{Kashikawa_SDF}(SDF;  {Kashikawa} {et~al.} 2004, hereafter
K04) in three
bands. Keck spectroscopy of over half of the SN hosts, and for several
hundred galaxies similar to the hosts, is used to estimate the
reliability of our photometric redshifts. The photometry of
the SNe and their hosts is used to classify the SNe with the algorithm
of P07.  As we
show below, we identify 33 SNe, classified as either Type Ia or core
collapse, with redshifts out to $\sim 1.7$, a number and redshift
range comparable to those in the {\it HST}/GOODS sample. While the
redshift distributions of the samples are consistent, there is a trend for
more high-$z$ SNe~Ia in the SDF. We derive the SN~Ia rate,
in four redshift bins, and find that while it is consistent, to within errors,
with previous results, at redshift $z\approx1.6$ it is about 50\% higher than
the value determined based on GOODS data, and at redshift
$z\approx0.7$ the rate we find is lower than that of GOODS.
We are currently in the process of increasing
our SN sample by obtaining additional epochs of the SDF, with the
objective of acquiring a sample of several hundred high-$z$ SNe.
Those observations and their analysis will be described in future
papers.

\section{Observations and Reductions}

\subsection{Imaging}

Our SN survey is carried out in the SDF ($\alpha = 13^h24^m39^s$,
$\delta = +27^\circ29'26''$; J2000). The reference images for this
field, obtained with the Suprime-Cam camera on the Subaru 8.2-m
telescope on Mauna Kea, Hawaii, are described by K04.
Suprime-Cam is a $5 \times 2$ mosaic of $2{\rm
k} \times 4{\rm k}$ CCDs at the prime focus of the telescope, with a
field of view of $ 34\arcmin \times 27\arcmin$, and a scale of
$0\farcs202~\textrm{pixel}^{-1}$.  This reference epoch of the SDF was
obtained by deep imaging, in five broad-band filters
($B,V,R,i',~\mathrm{and}~z'$) and two narrow-band filters ($NB816$ and
$NB921$), of an area of $30\arcmin \times 37\arcmin$, down to $3
\sigma$ limiting magnitudes of $B=28.45$, $V=27.74$, $R=27.80$,
$i'=27.43$, and $z'=26.62$ ($5
\sigma$ limits of $B=28.06$, $V=27.33$, $R=27.37$,
$i'=27.13$, and $z'=26.32$;
 \nocite{oke_gunn} here and throughout the paper, all
magnitudes are on the AB system;  {Oke} \& {Gunn} 1983). The imaging for the
reference SDF epoch was carried out between April 2002 and April 2003,
with some preliminary data collected in 2001, as described by K04.

We re-imaged the field on 2005 March 5 and 6 (UT dates are used
throughout this paper), in the three reddest Suprime-Cam broad bands:
$R$, $i'$, and $z'$. These bands were selected as being the most
efficient for finding high-redshift SNe, and for later classification
\nocite{POZ_TP1,AGY_TP2,RIESS_TP}(e.g.,  {Poznanski} {et~al.} 2002; {Gal-Yam} {et~al.} 2004b; {Riess} {et~al.} 2004b).  We obtained 22 exposures
of $360$~s each in $R$, 36 of $300$~s each in $i'$, and 76 of $240$~s
each in $z'$, for a total of $7920$~s ($R$), $10,800$~s ($i'$), and
$18,240$~s ($z'$). We followed a dithering pattern similar to the one
described by K04. Our seeing throughout the observations averaged
$0\farcs65$ in all three bands, ranging from $0\farcs5$ to $0\farcs8$.

We also observed the field with the MOSAIC-I wide-field imager on the
Kitt Peak National Observatory (KPNO) 4-m Mayall telescope, on 2005
March 2, April 5, and June 8, and on 2006 June 19.  The MOSAIC-I camera
consists of eight CCDs with a pixel size of $0\farcs258$
\nocite{Muller_98}({Muller} {et~al.} 1998), covering a field of view of $36\arcmin \times
36\arcmin$.  Each observation was in white light (i.e., unfiltered),
typically 1.5--2 hr long, split into dithered subexposures, with a
median seeing of about $0\farcs8$.  Compared to the $R$ band, the
unfiltered observations increase the throughput of stellar sources,
depending on their colours, by a factor of 2--4, and also increase the
level of the sky background by a factor of 4--8, depending on
conditions. The signal-to-noise ratio (S/N) is therefore multiplied by
$(2-4)\,/\,\sqrt{4-8} \approx (0.7-2)$.
Consequently, in many cases, unfiltered observations permit obtaining a
deeper image, closer to the depth reached with the larger-aperture Subaru telescope,
albeit with no colour
information. The 4-m data provide us with monitoring over longer times,
which is important for identifying variable
active galactic nuclei (AGNs), the main contaminant in our
SN search (see \S\ref{search}).

The {\it HST} archive includes Advanced Camera for Surveys (ACS)
images of a small portion of the SDF, which we use to examine the
morphology of one SN host galaxy that lies in this field. The data
were obtained with {\it HST}/ACS in 2002, on five separate occasions
between May 20 and July 3, in the F850LP band as part of program
GO-9075 (PI: S. Perlmutter).  The total exposure time is $5370$~s.

Subaru data were reduced following K04, using the Suprime-Cam pipeline
\textit{SDFRED} \nocite{YAGI_02,OUCHI_04}({Yagi} {et~al.} 2002; {Ouchi} {et~al.} 2004), and involving the following
main steps. The individual frames were overscan subtracted, flat
fielded using superflats, distortion corrected, matched to a common
point-spread function (PSF) of $0\farcs98$, sky subtracted,
registered, and combined. The combined image was then matched to the
reference image by using the
\textit{astrometrix}\footnote{\texttt{available at
http://www.na.astro.it/$\sim$radovich}} code to find the astrometric
correction, and the IRAF  \nocite{iraf}({Tody} 1986) task \textit{wregister} was used to register
the two images. Inducing PSF degradation to the new images may appear
questionable, since it reduces the frame depth and could adversely
affect the subsequent image subtraction. However, probably because our
reference epoch was reduced in such a manner, we found that the
optimal image subtraction, with the least numbers of artificial
residuals, was obtained by following the same procedure. The
photometric calibration of the first-epoch images was done by K04,
reaching a precision for the zero points of about 0.05 mag.  We
calibrated our images relative to the first epoch by comparing the
photometry of all the objects detected in both epochs.  The mean of
the differences between the two measurements was taken to be the
difference in zero points.  The final images reach $3\sigma$ limits
(defined as in K04) of $R=27.4$, $i'=27.1$, and $z'=26.3$
($5\sigma$ limits of $R=26.9$, $i'=26.7$), and $z'=26.0$),
shallower than the reference image by $0.5$~mag~($R$),
$0.4$~mag~($i'$), and $0.4$~mag~($z'$).

Reduction of the KPNO images was done using the IRAF package
\textit{MSCRED}, following the same steps as with the Subaru data,
except that PSF degradation was not applied.  The final images reach
$3\sigma$ limiting magnitudes corresponding approximately to
$R=25-26$~mag ($5\sigma$ limits of $R=24.5-25.3$~mag), depending on the epoch.
The depth reached implies
a S/N improvement in the filterless data, relative to $R$-band imaging,
of a factor near 2, for two of our epochs, and only a minor
gain for the remaining
two. {\it HST} images were reduced using the PyRAF script 
\textit{MultiDrizzle} \nocite{Koekemoer_02}({Koekemoer} {et~al.} 2002). 

We performed PSF matching and image subtraction between the new and
reference Subaru images in all bands, and between the four KPNO
epochs, using both the software ISIS \nocite{ALARD_98,ALARD_00}({Alard} \& {Lupton} 1998; {Alard} 2000) and the
Common PSF Method (CPM; \nocite{AGY_03lw}{Gal-Yam} {et~al.} 2004a; Gal-Yam et al. 2007, in preparation).
Briefly, ISIS minimizes a spatially variable convolution
kernel that degrades the PSF of the better-seeing image to the PSF of
the worse-seeing image.  CPM convolves each image with the measured
PSF of the other, and thus the final PSF of both images is worse than
the initial PSF of either image, but kernel-matching, and its inherent
uncertainties in the presence of noise and pixellation, are avoided.
Both procedures produce output images with nominally identical PSFs,
while introducing a minimal amount of noise. While ISIS is robust and
stable, we find that in some cases it tends to produce more
subtraction artefacts than CPM, mimicking SNe near the cores of bright
galaxies. We have therefore used both codes.

As a consequence of the dithering, the final images have a field of
view of $0.31~\mathrm{deg^2}$, but with a substantial region along the
edges where the S/N is significantly
smaller. This is due to the different effective exposure on the
fringes of the field.  We therefore trim down the difference image to
81\% of its full size, and remain with a total differenced area of
$0.25~\mathrm{deg^2}$ (which is, in fact, the size of the Suprime-Cam
field).

\subsection{Spectroscopy}

We obtained spectra of some of our SN host galaxies (see
\S\ref{hosts}), and of several hundred random galaxies in the
SDF. Observations were carried out on 2007 January 12, January 22,
March 16, and April 12, using the Low-Resolution Imaging Spectrometer
\nocite{oke_95}(LRIS;  {Oke} {et~al.} 1995) on the Keck~I telescope 
for the first night,
and the Deep Imaging Multi-Object Spectrograph \nocite{Faber_03}(DEIMOS;
 {Faber} {et~al.} 2003) on the Keck II telescope for the three other
nights. Eight unique object masks and positions where used, with up to
12 host galaxies per mask (including also targets from Subaru
observations in February 2007, which are not analysed in this
paper), and several tens of random galaxies per mask. Each mask was
typically observed for $3\times1800$~s.

The LRIS mask was observed with the 600 line mm$^{-1}$ grism blazed
at 4000~\AA, and the 400 line mm$^{-1}$ grating blazed at 8500~\AA,
together with the D560 dichroic. This typically yields a wavelength 
range of $\sim$3000--10,000~\AA, but the specific position of the slit
on the mask shifts the wavelength range for each individual spectrum.
The spectra have resolutions of $\sim$3.5~\AA\ and $\sim$7.0~\AA\ for
the blue and red sides, respectively.

All but one of the DEIMOS masks were observed with the 600 line mm$^{-1}$
grating blazed at 7500~\AA, and the GG495 order-blocking filter.  We chose a
wavelength range of $\sim$5000--10,000~\AA, with the precise limits depending on
each individual spectrum (different slit positions).  One mask was designed to
target $1.5 < z < 2$ galaxies as well as slightly higher-redshift SN hosts (see
\S\ref{hosts:z}).  For this mask, we observed with the 600 line mm$^{-1}$
grating and the OG550 order-blocking filter, with a central wavelength of about
8000~\AA. This yielded a typical wavelength range of 5500--10,500~\AA, without
second-order contamination at longer wavelengths, due to the different filter.

The 600 line mm$^{-1}$ grating yields a full width at half-maximum
intensity (FWHM) resolution of $\sim$1.5~\AA, or
$\sim$150~km~s$^{-1}$, at 7500~\AA.  This resolution is sufficient to
resolve many night-sky lines and the [\mbox{O\,{\sc ii}}] $\lambda\lambda$3726,
3729 doublet. Resolving night-sky lines is useful for finding emission
lines in the reddest part of the spectrum, where sky lines are blended
in low-resolution spectra.  Resolving the [\mbox{O\,{\sc ii}}] doublet
allows us to confidently identify an object's redshift even with a
``single'' line.

The DEIMOS data were reduced using a modified version of the DEEP2 data
reduction
pipeline\footnote{http://astro.berkeley.edu/$\sim$cooper/deep/spec2d/},
which bias corrects, flattens, rectifies, and sky subtracts the data
before extracting a spectrum \nocite{foley_07}({Foley} {et~al.} 2007).
The LRIS data were  reduced using a combination of typical IRAF 
techniques and our own IDL procedures \nocite{foley_07}({Foley} {et~al.} 2007). The wavelength
solutions were derived by low-order polynomial fits to the lamp spectral
lines, and shifted to match night-sky lines at the positions of the
objects.  Standard-star spectra were obtained through a long slit on
the same night, and were used to flux calibrate the spectra and remove
telluric absorption \nocite{matheson_01}({Matheson} {et~al.} 2001).  To obtain the proper
absolute flux scale and to correct for minor continuum differences due
to slit losses, the spectra were scaled to match the $R$-band and
$i'$-band photometry of the galaxies.

\section{Supernova Candidates}\label{search}
\subsection{Candidate Selection}

The $z'$-band difference image obtained using ISIS was scanned by eye
to search for variable and transient objects in the new image. Morokuma et al. (in
preparation) will present a study of the various variable objects
they detect within the subframes that constitute our
reference image set. They have also identified nearly a
thousand AGNs in the SDF based on
their long-term $i'$-band variability.
In our search, these galaxies were therefore
ignored (clearly non-nuclear variable sources in these objects were
searched for, but none was found).  The remaining variable candidates
were examined as follows, in order to reject other non-SNe.

\begin{enumerate}
\item We compared the $z'$-band difference images relative to the
reference SDF epoch for the first and second nights of the data
separately, to identify and reject moving objects (asteroids) and
possible subtraction artefacts and noise peaks.
\item ISIS and CPM subtraction images were compared to test for
possible subtraction artefacts, especially near bright
galaxies. Candidates that had distorted or otherwise suspicious shapes
in the ISIS subtraction image, and were clearly absent from the CPM
image, were rejected.
\item KPNO subtraction images were used to identify AGNs based on
their non-SN-like (e.g., slowly rising) light curves. For 10 of our
candidates, at the bright end of the distribution, we have detections
also in one or more of the KPNO images, with a photometric behavior
consistent with SNe, i.e., declining on time scales of a month or two.
\item For every candidate found in the $z'$ band, subtraction images
in the $R$ and $i'$ bands were also examined, and objects which showed
suspect residual shapes, indicative of a subtraction artefact, were
rejected. We note that no candidate was rejected because of a
non-detection in one band or another, since at least some high-$z$ SNe
are expected to be very faint or undetected in the observed-frame $R$
and $i'$ bands.
\end{enumerate}
We further note that the purity of our SN sample will increase retroactively
as we obtain additional epochs of this field, and identify any
remaining AGNs that may still contaminate the sample.

\subsection{Detection Efficiency Simulations}\label{s:eff}

In order to assess our detection efficiency, artificial point sources,
with characteristics matching as closely as possible those of the SN
population, were planted blindly in the $z'$-band image and were
searched for, together with the real transients. The simulated sample
was constructed as follows. First, we calculated the photometric
redshifts of the entire galaxy sample in the SDF (about 150,000
objects; see K04), using the Bayesian photometric redshift (photo-$z$)
code BPZ \nocite{Benitez_00}({Ben{\'{\i}}tez} 2000). Briefly, BPZ fits a set of template
galaxies to the photometry, with the possibility to use a prior on the
redshift.  We used the same 6 template galaxies as \nocite{Benitez_00}{Ben{\'{\i}}tez} (2000),
from \nocite{CWW_80}{Coleman}, {Wu} \& {Weedman} (1980) and \nocite{Kinney_96}{Kinney} {et~al.} (1996).  As a prior we took the
redshift distribution of galaxies in the Hubble Deep Field, which has
a depth comparable to that of the SDF \nocite{fernandez-soto_99}({Fern{\'a}ndez-Soto}, {Lanzetta} \& {Yahil} 1999).
The errors given by BPZ were used to select a subsample with
relatively robust redshifts (i.e., small nominal uncertainties), while
requiring the selected galaxies to have a redshift distribution
similar to the one found in the Hubble Deep Field.
A discussion of the photo-$z$ determination for the actual SN host galaxies,
along with our spectroscopic training set, are presented in \S\ref{hosts}.

In order to account for the population of galaxies which are below our
limiting magnitude, but can still host SNe, we extrapolated, in every
redshift bin, the galaxy population using a \nocite{schechter_76}{Schechter} (1976)
luminosity function and assigned those extrapolated galaxies random
positions within our field. SN rates are expected to roughly follow
the luminosities of the galaxies. We approximated the luminosity of
each galaxy using its magnitude and luminosity distance (assuming
here, and throughout the paper, cosmological parameters
$H_0=70~\mathrm{km~s}^{-1}~\mathrm{Mpc}^{-1}$, $\Omega_m=0.3$,
$\Omega_\Lambda=0.7$).  These luminosities were used to weight each
galaxy's probability to host a SN.  The position of the simulated SNe
within each host galaxy was again set to follow the light. We defined
for each galaxy, using SExtractor \nocite{Bertin_96}({Bertin} \& {Arnouts} 1996), 10 annuli, each
containing one-tenth of the total flux of the galaxy. The radial
position of the simulated SNe was randomly drawn to lie inside one of
those 10 annuli. While the annuli are circular, the hosts are
generally not.  To approximately compensate for this, the position
angles of the SNe, relative to the galaxies' major axes, were chosen
from a linear distribution, where the major axis position angle has
the highest probability, and the orthogonal angle the lowest.

\begin{figure}
\center
\includegraphics[width=0.45\textwidth]{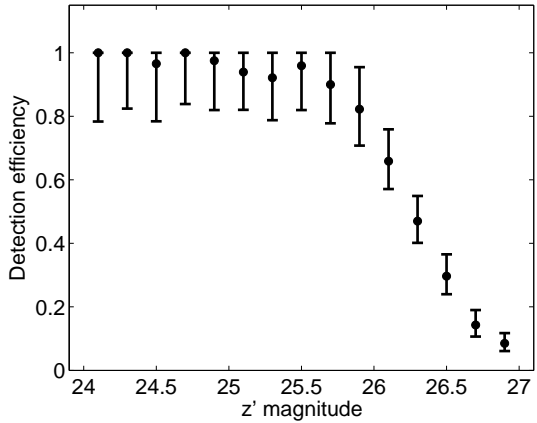}
\caption{Fraction of simulated SNe recovered as a function of $z'$-band
magnitude. Error bars indicate $1\sigma$ Poisson uncertainties.\label{f:eff}}
\end{figure}

Each fake SN was assigned a random magnitude, but was kept in the
sample only if it was no brighter than the magnitude a SN Ia would
reach at peak at the given redshift of the host. We thus generated a
list of positions and magnitudes for more than a thousand fake
SNe. The objects were planted in the image using the IRAF task
\textit{mkobject} with a Gaussian profile.
During the search, we could not distinguish between
real and fake SNe. Our resulting efficiency as a function of magnitude
can be seen in Figure \ref{f:eff}. Our recovery rate is stable and
nearly perfect, averaging 96\%, up to $z'=25$ mag, where it starts to
decline. The efficiency is 50\% at $z'=26.3$ mag, and effectively
reaches zero at $z'=27$ mag.  More than 80\% of the 1000 simulated SNe
are in the interesting magnitude range, between 25 and 27, where a
good sampling of the efficiency curve is important.
We have tested the effect of replacing the Gaussian profile of the fake SNe
by an empirical profile measured from the image, using different template stars
for different portions of the image, and found no changes in the efficiency compared
to our previous simulations.

\begin{figure*}
\center
\includegraphics[width=0.8\textwidth]{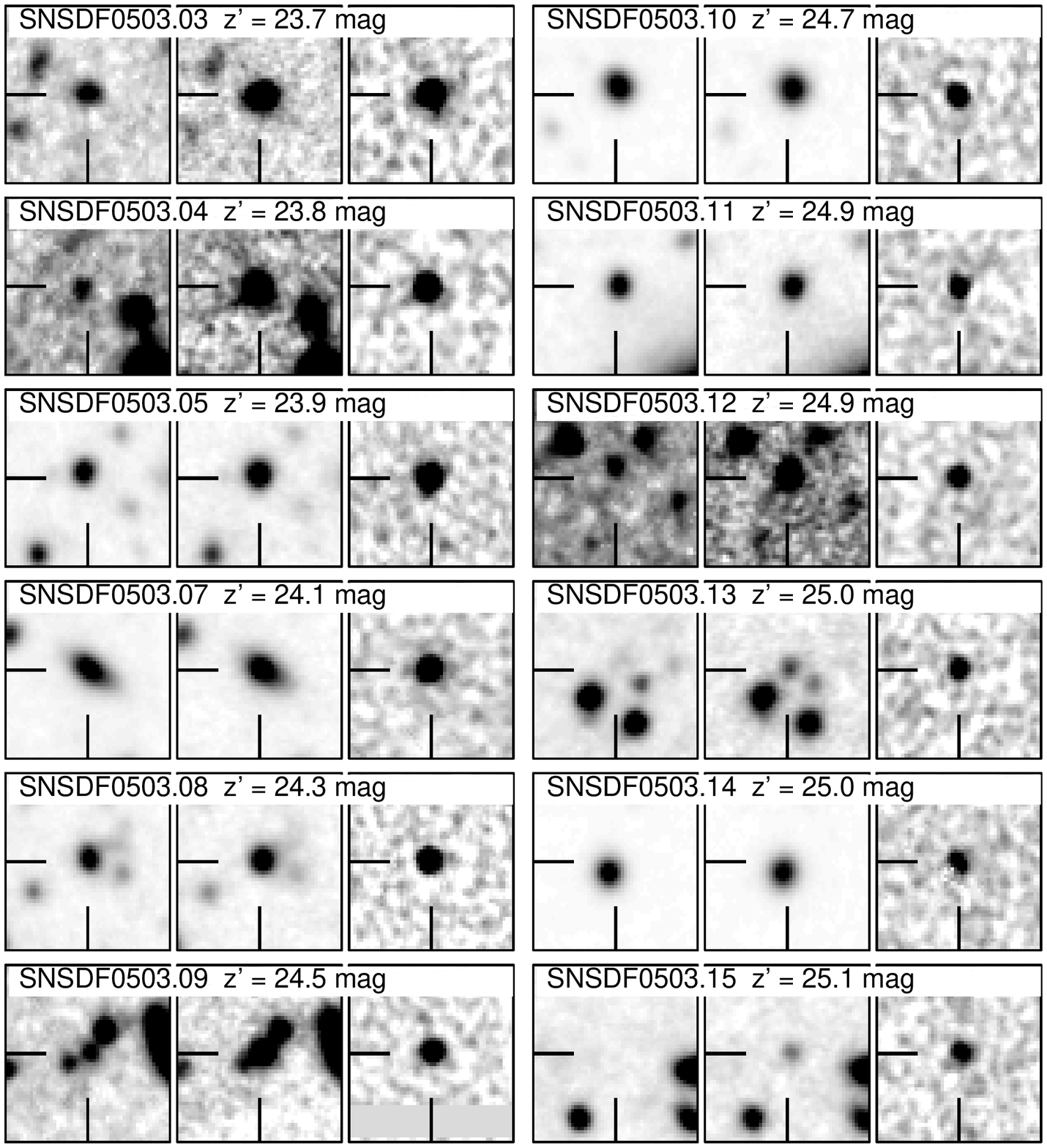}
\end{figure*}
\begin{figure*}
\center
\includegraphics[width=0.8\textwidth]{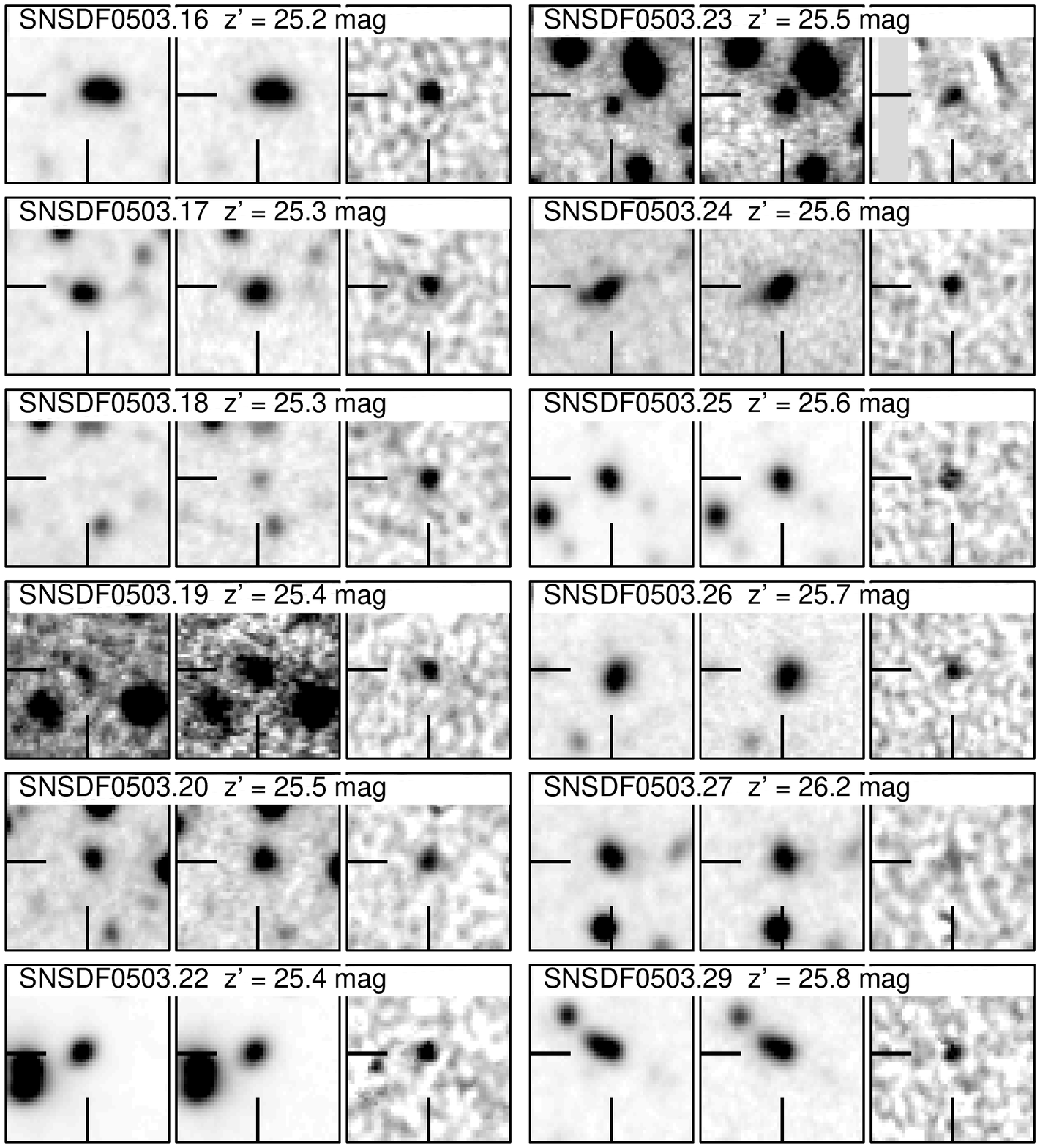}
\end{figure*}
\begin{figure*}
\center
\includegraphics[width=0.8\textwidth]{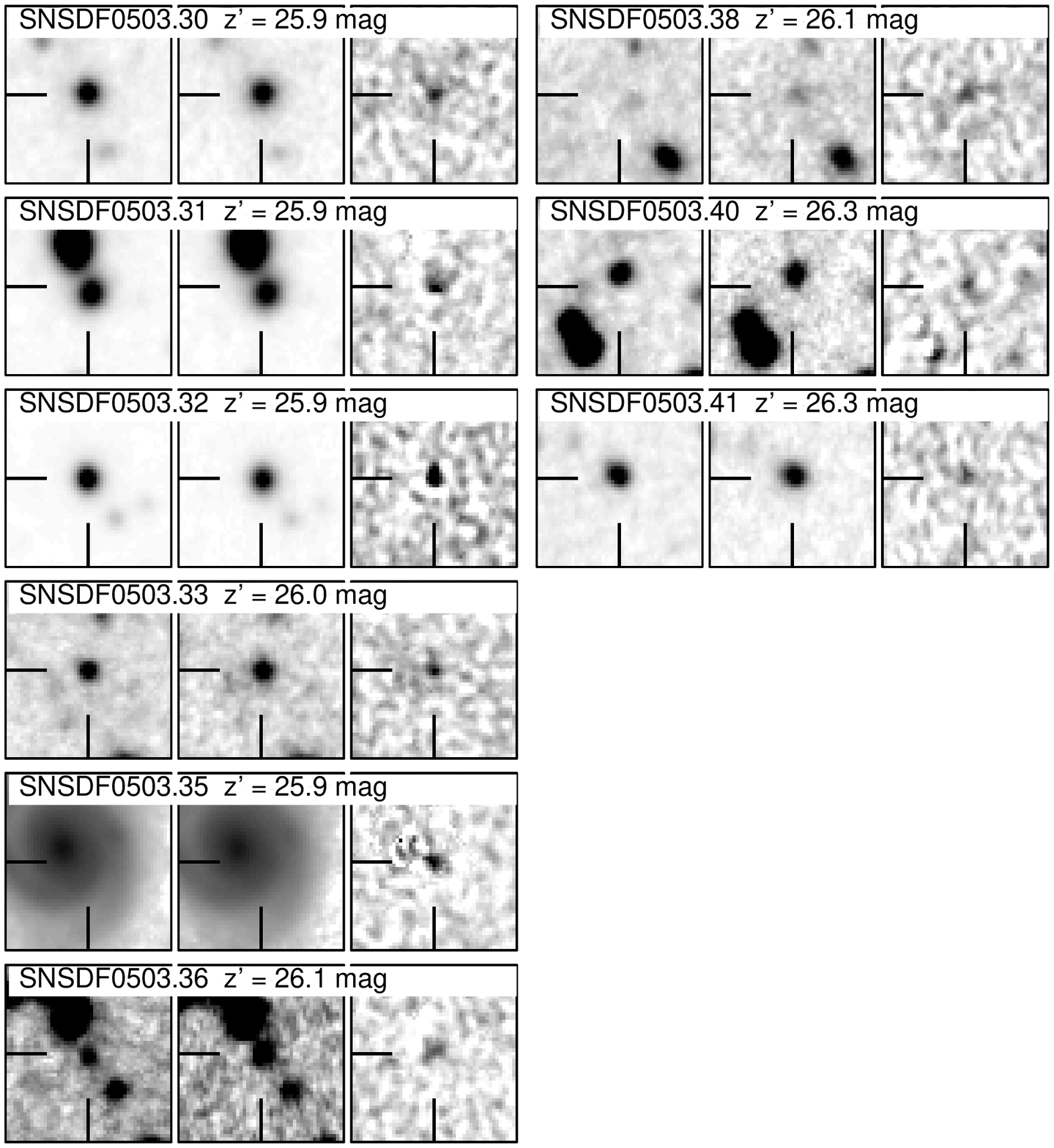}
\caption{Image sections, $10''$ on a side, for the SNe in our
sample. North is up, east is to the left.  For each SN we show the reference
and new epoch $z'$-band images (left and centre, respectively), and
the difference image (right).  The greyscale for all the difference
images is the same, while for the other images it is chosen to
emphasize host-galaxy details.\label{f:stamps}}
\end{figure*}

\subsection{Supernova Sample}\label{sample}

We found a total of 33 SNe, shown in Figure \ref{f:stamps}, with
magnitudes in the range $z'=23.7$ to $z'=26.3$. Table \ref{t:SNe}
lists the SNe and their properties.
Our SNe do not have
spectroscopic confirmation (indeed, spectroscopy for most of them is
impossible with existing telescopes) or light curves, therefore they do not
satisfy the International Astronomical Union criteria for a
``standard" appellation.  Nonetheless, since we are confident in our
search and screening process (and also for the sake of brevity), we
will refer to them as SNe, rather than SN candidates. We denote the
SNe from this March 2005 run as ``SNSDF$0503.$XX", XX being a serial
number, ordered roughly according to the SN $z'$-band magnitude.  The
respective host galaxies are referred to as ``hSDF$0503.$XX".  Ten of
the 11 brightest SNe are also visible in the first image from KPNO.
Apart from the 33 SNe, we detect several tens of candidates at fainter
magnitudes, as we would expect based on our efficiency simulations
(see \ref{s:eff}), but these are 1$\sigma$ to 2$\sigma$ detections,
which are mixed with an unknown number of false positives that arise
from random peaks in the noise distribution and from subtraction
artefacts. We therefore limit our sample to $z'\leq26.3$ mag,
which leaves only candidates that we judge to be unambiguously real.
Coincidentally, $z'\leq26.3$ mag is also our 50\% detection efficiency limit.

\begin{table*}
\scriptsize
\begin{minipage}{\textwidth}
\begin{tabular}{l|c|c|c|c|c|c|c|c|c|c|c|c|c}
\hline
\hline
{ID}  & {$\alpha$ (J2000)} & {$\delta$ (J2000)} & {$R$} & {$i'$} & {$z'$} & S/N$^a$ & {Photo-$z$} & {Spec-$z$} & {$P_{Ia}$$^b$} & {Type} & {Posterior $z$}
&  {$\chi^2$}  &M$^c0$\\
\hline
SNSDF0503.03  &  13:23:52.42  &  +27:12:45.36  &  24.58$\pm$0.08  &  23.98$\pm$0.08  &  23.71$\pm$0.08  &  35  &  0.91     & 0.886     & 0.98  & Ia  & 0.89  & 0.95   & $-19.3$ ($B$) \\
SNSDF0503.04  &  13:24:45.54  &  +27:18:14.06  &  24.08$\pm$0.08  &  24.04$\pm$0.08  &  23.83$\pm$0.08  &  33  &  0.43     &  \nodata  & 0.32  & CC  & 0.34  & 0.08   & $-17.3$ ($B$) \\
SNSDF0503.05  &  13:24:22.02  &  +27:16:06.99  &  24.70$\pm$0.08  &  24.25$\pm$0.08  &  23.85$\pm$0.08  &  56  &  0.59     & 0.593     & 0.90  & Ia  & 0.59  & 1.19   & $-17.8$ ($B$) \\
SNSDF0503.07  &  13:25:14.55  &  +27:29:16.46  &  25.26$\pm$0.08  &  24.52$\pm$0.08  &  24.12$\pm$0.08  &  37  &  0.91     & 0.918     & 0.95  & Ia  & 0.92  & 0.93   & $-19.0$ ($B$) \\
SNSDF0503.08  &  13:25:33.35  &  +27:36:39.61  &  25.10$\pm$0.08  &  24.51$\pm$0.08  &  24.26$\pm$0.08  &  46  &  0.69     & 0.707     & 0.78  & Ia  & 0.71  & 0.20   & $-18.0$ ($B$) \\
SNSDF0503.09  &  13:24:37.91  &  +27:36:38.00  &  25.64$\pm$0.09  &  24.99$\pm$0.08  &  24.55$\pm$0.09  &  27  &  0.67     &  \nodata  & 0.61  & Ia  & 0.67  & 0.02   & $-17.4$ ($B$) \\
SNSDF0503.10  &  13:25:28.58  &  +27:36:24.60  &  25.96$\pm$0.11  &  24.78$\pm$0.08  &  24.74$\pm$0.11  &  22  &  0.81     & 0.849     & 0.89  & Ia  & 0.85  & 5.34   & $-18.1$ ($B$) \\
SNSDF0503.11  &  13:25:06.01  &  +27:40:22.31  &  24.90$\pm$0.08  &  24.91$\pm$0.08  &  24.94$\pm$0.12  &  11  &  0.57     &  \nodata  & 0.50  & Ia  & 0.57  & 2.32   & $-17.2$ ($B$) \\
SNSDF0503.12  &  13:24:00.48  &  +27:26:04.13  &  27.08$\pm$0.31  &  25.40$\pm$0.10  &  24.93$\pm$0.12  &  17  &  0.73     &  \nodata  & 0.62  & Ia  & 1.44  & 0.61   & $-17.1$ ($U$) \\
SNSDF0503.13  &  13:23:54.85  &  +27:34:17.23  &  25.23$\pm$0.08  &  25.02$\pm$0.08  &  25.02$\pm$0.13  &  10  &  0.58     &  \nodata  & 0.17  & CC  & 0.51  & 0.04   & $-17.0$ ($B$) \\
SNSDF0503.14  &  13:24:09.34  &  +27:18:41.91  &  25.69$\pm$0.10  &  25.10$\pm$0.09  &  25.04$\pm$0.13  &  11  &  0.53     & 0.506     & 0.27  & CC  & 0.51  & 0.39   & $-16.3$ ($B$) \\
SNSDF0503.15  &  13:24:08.09  &  +27:35:21.78  &  $>27.10$$^d$    &  26.05$\pm$0.17  &  25.08$\pm$0.14  &  13  &  \nodata  &  \nodata  & 0.71  & Ia  & 1.27  & 0.00   & $-19.4$ ($B$) \\
SNSDF0503.16  &  13:24:40.09  &  +27:18:34.25  &  25.97$\pm$0.12  &  25.52$\pm$0.11  &  25.18$\pm$0.15  &  13  &  1.15     &  \nodata  & 0.87  & Ia  & 1.08  & 1.79   & $-18.7$ ($B$) \\
SNSDF0503.17  &  13:25:06.12  &  +27:22:32.44  &  $>27.10$$^d$    &  26.75$\pm$0.33  &  25.32$\pm$0.17  &  18  &  1.36     &  \nodata  & 0.88  & Ia  & 1.36  & 1.07   & $-19.5$ ($B$) \\
SNSDF0503.18  &  13:25:14.35  &  +27:28:52.83  &  $>27.10$$^d$    &  26.02$\pm$0.17  &  25.31$\pm$0.17  &   9  &  \nodata  &  \nodata  & 0.57  & Ia  & 1.60  & 0.47   & $-18.4$ ($U$) \\
SNSDF0503.19  &  13:24:50.37  &  +27:45:16.61  &  $>27.10$$^d$    &  26.46$\pm$0.25  &  25.44$\pm$0.19  &   8  &  1.89     &  \nodata  & 0.70  & Ia  & 1.68  & 0.51   & $-15.6$ ($U$) \\
SNSDF0503.20  &  13:24:57.75  &  +27:36:41.80  &  $>27.10$$^d$    &  26.63$\pm$0.30  &  25.48$\pm$0.20  &  10  &  0.81     &  \nodata  & 0.80  & Ia  & 0.88  & 0.06   & $-16.8$ ($B$) \\
SNSDF0503.22  &  13:24:35.31  &  +27:19:41.64  &  25.75$\pm$0.10  &  25.42$\pm$0.10  &  25.43$\pm$0.19  &   7  &  0.43     & 0.450     & 0.17  & CC  & 0.45  & 1.17   & $-15.9$ ($B$) \\
SNSDF0503.23  &  13:24:48.19  &  +27:45:27.67  &  $>27.10$$^d$    &  26.31$\pm$0.22  &  25.54$\pm$0.21  &   7  &  0.47     &  \nodata  & 0.25  & CC  & 0.46  & 0.58   & $-14.3$ ($B$) \\
SNSDF0503.24  &  13:24:21.80  &  +27:31:41.92  &  26.54$\pm$0.19  &  25.23$\pm$0.09  &  25.56$\pm$0.21  &  10  &  0.19     & 0.195     & 0.00  & CC  & 0.20  & 22.17  & $-13.2$ ($V$) \\
SNSDF0503.25  &  13:24:21.78  &  +27:13:22.77  &  $>27.10$$^d$    &  26.28$\pm$0.21  &  25.63$\pm$0.22  &   9  &  0.51     & 0.530     & 0.36  & CC  & 0.53  & 0.62   & $-15.1$ ($B$) \\
SNSDF0503.26  &  13:24:21.51  &  +27:41:10.51  &  $>27.10$$^d$    &  26.74$\pm$0.33  &  25.71$\pm$0.24  &   7  &  0.81     & 1.130     & 0.81  & Ia  & 1.13  & 0.52   & $-18.2$ ($B$) \\
SNSDF0503.27  &  13:25:24.24  &  +27:35:46.68  &  $>27.10$$^d$    &  26.72$\pm$0.32  &  26.19$\pm$0.37  &   5  &  0.84     & 0.848     & 0.52  & Ia  & 0.85  & 0.10   & $-16.5$ ($B$) \\
SNSDF0503.29  &  13:24:42.38  &  +27:14:23.40  &  26.43$\pm$0.17  &  26.08$\pm$0.18  &  25.84$\pm$0.27  &   4  &  0.90     & 1.085     & 0.80  & Ia  & 1.08  & 2.82   & $-17.8$ ($B$) \\
SNSDF0503.30  &  13:24:12.86  &  +27:37:47.55  &  $>27.10$$^d$    &  26.93$\pm$0.39  &  25.89$\pm$0.29  &   8  &  1.31     &  \nodata  & 0.72  & Ia  & 1.30  & 0.52   & $-18.7$ ($B$) \\
SNSDF0503.31  &  13:24:28.67  &  +27:44:47.68  &  26.29$\pm$0.15  &  26.01$\pm$0.17  &  25.88$\pm$0.28  &   6  &  0.96     & 1.080     & 0.68  & Ia  & 1.08  & 4.87   & $-17.7$ ($B$) \\
SNSDF0503.32  &  13:25:27.90  &  +27:33:06.09  &  $>27.10$$^d$    &  $>27.00$$^d$    &  25.88$\pm$0.28  &   7  &  1.05     & 1.214     & 0.82  & Ia  & 1.21  & 1.03   & $-18.6$ ($B$) \\
SNSDF0503.33  &  13:24:02.89  &  +27:17:53.77  &  26.57$\pm$0.19  &  26.70$\pm$0.32  &  26.03$\pm$0.32  &   7  &  1.22     & \nodata   & 0.72  & Ia  & 0.86  & 4.33   & $-17.7$ ($B$) \\
SNSDF0503.35  &  13:24:22.24  &  +27:15:14.85  &  $>27.10$$^d$    &  25.54$\pm$0.11  &  25.93$\pm$0.30  &   7  &  0.34     & 0.340     & 0.01  & CC  & 0.34  & 30.07  & $-13.2$ ($V$) \\
SNSDF0503.36  &  13:24:05.12  &  +27:38:45.62  &  26.75$\pm$0.23  &  26.34$\pm$0.23  &  26.14$\pm$0.36  &   7  &  0.87     &  \nodata  & 0.46  & CC  & 0.87  & 0.26   & $-16.8$ ($B$) \\
SNSDF0503.38  &  13:24:17.97  &  +27:15:43.49  &  $>27.10$$^d$    &  $>27.00$$^d$    &  26.12$\pm$0.35  &   5  &  1.87     &  \nodata  & 0.41  & CC  & 0.30  & 0.43   & $-17.0$ ($V$) \\
SNSDF0503.40  &  13:23:39.00  &  +27:21:03.11  &  $>27.10$$^d$    &  $>27.00$$^d$    &  26.34$\pm$0.43  &   3  &  1.83     & 1.564     & 0.95  & Ia  & 1.56  & 0.96   & $-15.1$ ($U$) \\
SNSDF0503.41  &  13:25:22.36  &  +27:41:02.41  &  $>27.10$$^d$    &  26.43$\pm$0.25  &  26.34$\pm$0.42  &   4  &  0.70     & 0.709     & 0.32  & CC  & 0.71  & 1.51   & $-15.9$ ($B$) \\

\hline
\end{tabular}
\end{minipage}
\caption{Supernovae Discovered in the Subaru Deep Field.
$^a$ Signal-to-noise ratio of the SN, in the $z'$-band image.
$^b$ Probability that object is a SN~Ia, based on SN-ABC (P07).
$^c$ Approximate absolute magnitudes in rest-frame $B$ band, unless rest-frame $B$ does not
fall within the observed $R, i', z'$ bands, in which case we list rest-frame $U$ or $V$ absolute magnitudes, at high or low redshift, respectively.
$^d$ Fainter than $3\sigma$ limiting magnitude in this band.}
\label{t:SNe}
\end{table*}


Aperture photometry of the SNe was performed on the difference $R$,
$i'$, and $z'$ images, using SExtractor, and adopting a $2\arcsec$
diameter circular aperture.  To estimate the photometric uncertainty,
we measured the magnitudes of the $1000$ simulated transients planted
in the image for the efficiency measurements (see \S\ref{s:eff}), and
took the root mean square (rms) in each magnitude bin to be the
minimum photometric error for objects of that magnitude. The adopted
uncertainty for each SN was taken to be the larger among the formal
error computed by SExtractor and the statistical error for the given
magnitude from the simulated objects.
For each SN we derived the local S/N by comparing the SN counts in the
photometric aperture, to the standard deviation of the total counts in tens of
identical apertures on nearby, similar, background.
We also measure the offset
between the simulated SN positions as input to the images, and as
found by SExtractor. All the simulated SNe are found within
$0\farcs10$ of their intended positions, and 93\% are within
$0\farcs04$.  We use these results to estimate the accuracy to which an
offset between a SN and its host-galaxy centre can be detected.

The two SNe closest to their respective host centres have measured
offset of $0\farcs07$ and $0\farcs08$, while all others have offsets
greater than $0\farcs10$, effectively ruling out an AGN classification
for most, if not all, objects.

\section{Supernova Host Galaxies}\label{hosts}

In this section, we determine which galaxy hosts every SN, and
measure its properties. This allows us to measure the redshift of the SNe,
and eventually, will permit a study of the correlations between the 
properties of SNe and their host galaxies.

\subsection{Identification and Photometry}

For our list of SNe, we compiled a list of potential host galaxies. We
measured the Petrosian magnitudes \nocite{petrosian_76}({Petrosian} 1976) of the host
galaxies on the reference (deeper) epoch, in seven bands, with
SExtractor. Petrosian magnitudes allow one to measure the flux of resolved
objects within a given fraction of the light profile, without the
dependence on the amplitude of the surface brightness profile associated
with ``isophotal'' magnitudes \nocite{blanton_01,graham_05}(e.g.,  {Blanton} {et~al.} 2001; {Graham} {et~al.} 2005).
As for the SN photometry, we estimated the uncertainty in
each magnitude bin using artificial sources with galactic profiles
that we planted in the images.

The hosts of the SNe were chosen based on the smallest separations
between each SN and its neighboring galaxies, in terms of each of the
galaxies' half-light radii, as measured with SExtractor in the $i'$
band. For most of the SNe, the choice was obvious, since the SNe and
their hosts were unresolved from each other.

For two SNe (SNSDF$0503.15$ and SNSDF$0503.18$) we found no plausible host
within 6 half-light radii.  Assuming a \nocite{Sersic_68}{Sersic} (1968) model for the
galaxy radial profile between $n=4$ \nocite{de_Vaucouleurs_48,peng_02}(de Vaucouleurs law;  {de Vaucouleurs} 1948; {Peng} {et~al.} 2002)
and $n=1$ \nocite{freeman_70,peng_02}(exponential disk;  {Freeman} 1970; {Peng} {et~al.} 2002),
we find that between 91\% and 99.99\% of the light,
respectively, is within 6 half-light radii . The probability that a SN
belongs to a host at a distance beyond this limit is thus small.  Due
to the high density of sources in the SDF, using a larger limit would
lead to confusion regarding the hosts of most of the SNe, including
those SNe which are unresolved from a particular galaxy.  The two SNe
which we label as ``hostless'' are probably the result of a host that
is below our detection threshold in all the photometric bands
\nocite{AGY_IGSN}(though these SNe could, in fact, be hostless; see
 {Gal-Yam} {et~al.} 2003).  Alternative explanations are that these candidates are
actually flaring Galactic M-dwarfs or variable AGNs in faint hosts.
The non-detection of a source at this position in the reference epoch,
down to $i'\approx 27.5$~mag, means that a star with a quiescent absolute
magnitude of $M_i \approx 8$ \nocite{the_84}(e.g.,  {Th\'{e}}, {Steenman} \& {Alcaino} 1984) would be at a distance
$>80$~kpc, making this option highly improbable.  The possibility of a
high-$z$ AGN cannot be excluded.  Nevertheless, the $\chi^2$ criterion
we apply to the candidates (see \S\ref{class}) does not support this
interpretation.  Table \ref{t:hosts} lists the SN hosts and their
properties.

\begin{table*}
\vspace{4cm}
\vbox to220mm{\vfil
\includegraphics[width=1\textwidth,angle=90]{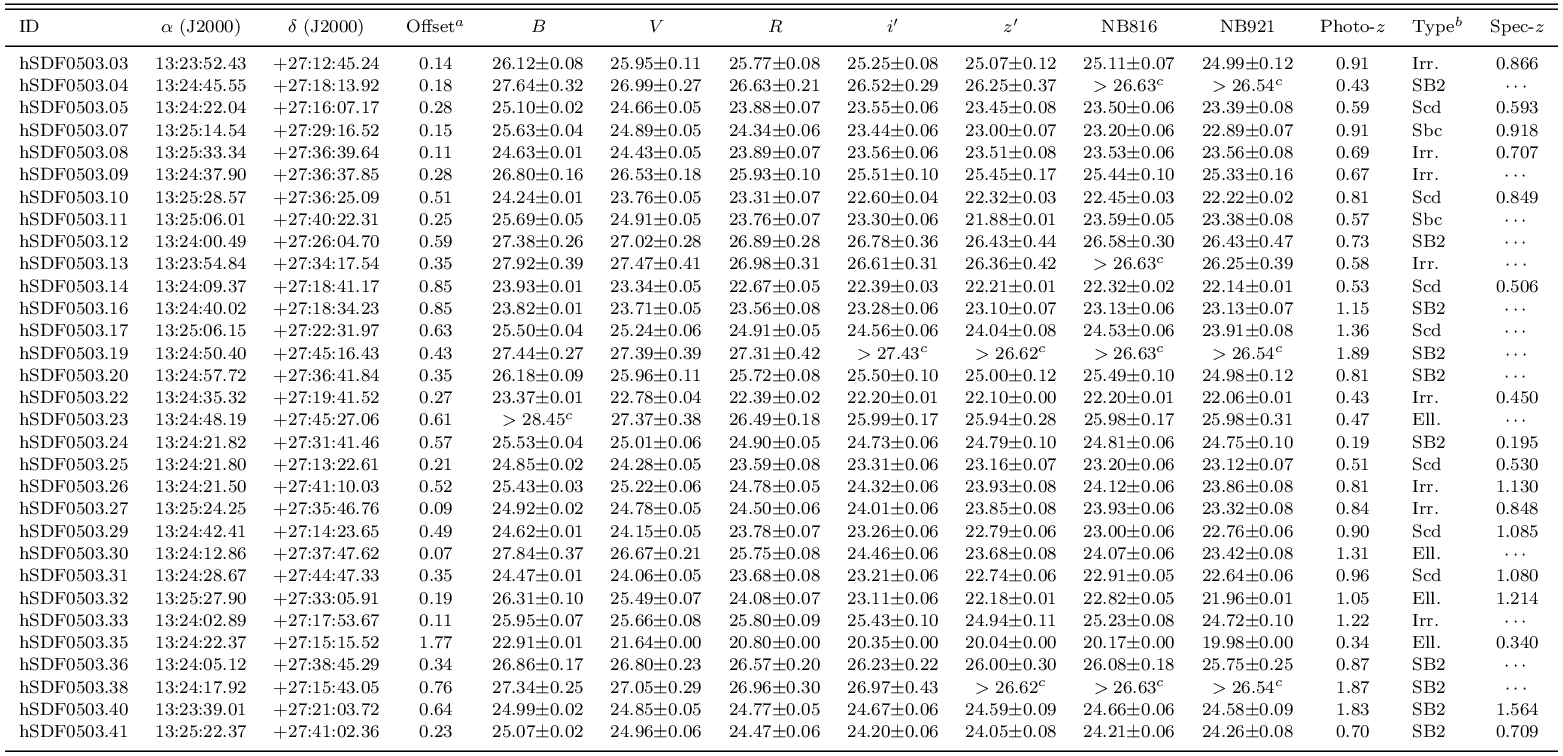}

\vspace{-1cm}
 \caption{SN Host Galaxies. $^a$ Arcseconds.
 $^b$ Best-fitting galaxy template. SB2 is a starburst spectral template from
 Kinney et al. (1996).
 $^c$ Fainter than $3\sigma$ limiting magnitude in this band.\label{t:hosts}}
\vfil}
\end{table*}

\subsection{Host Redshifts}\label{hosts:z}

\begin{figure*}
\center
\includegraphics[width=0.9\textwidth]{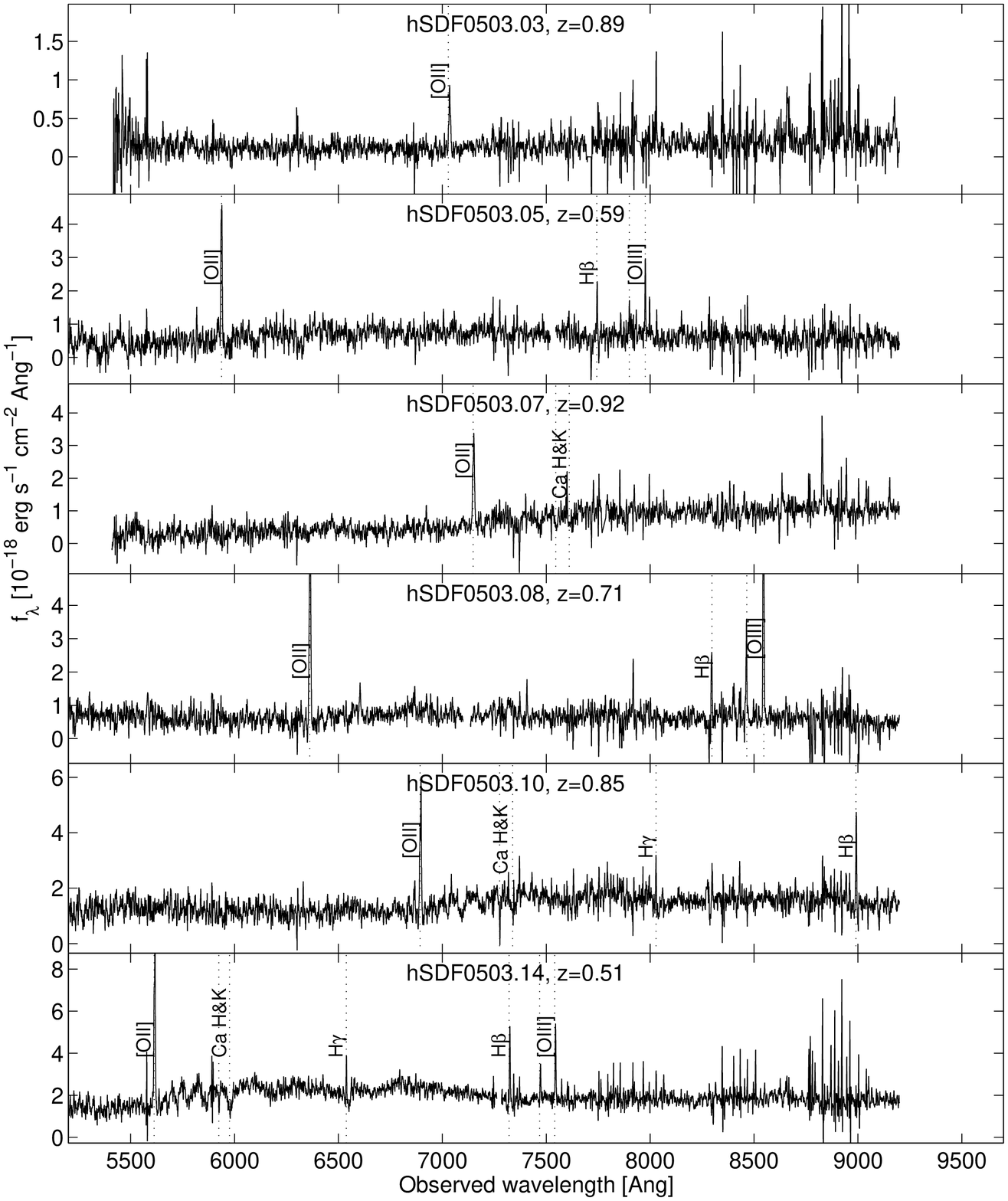}
 \end{figure*}
 \begin{figure*}
\center
\includegraphics[width=0.9\textwidth]{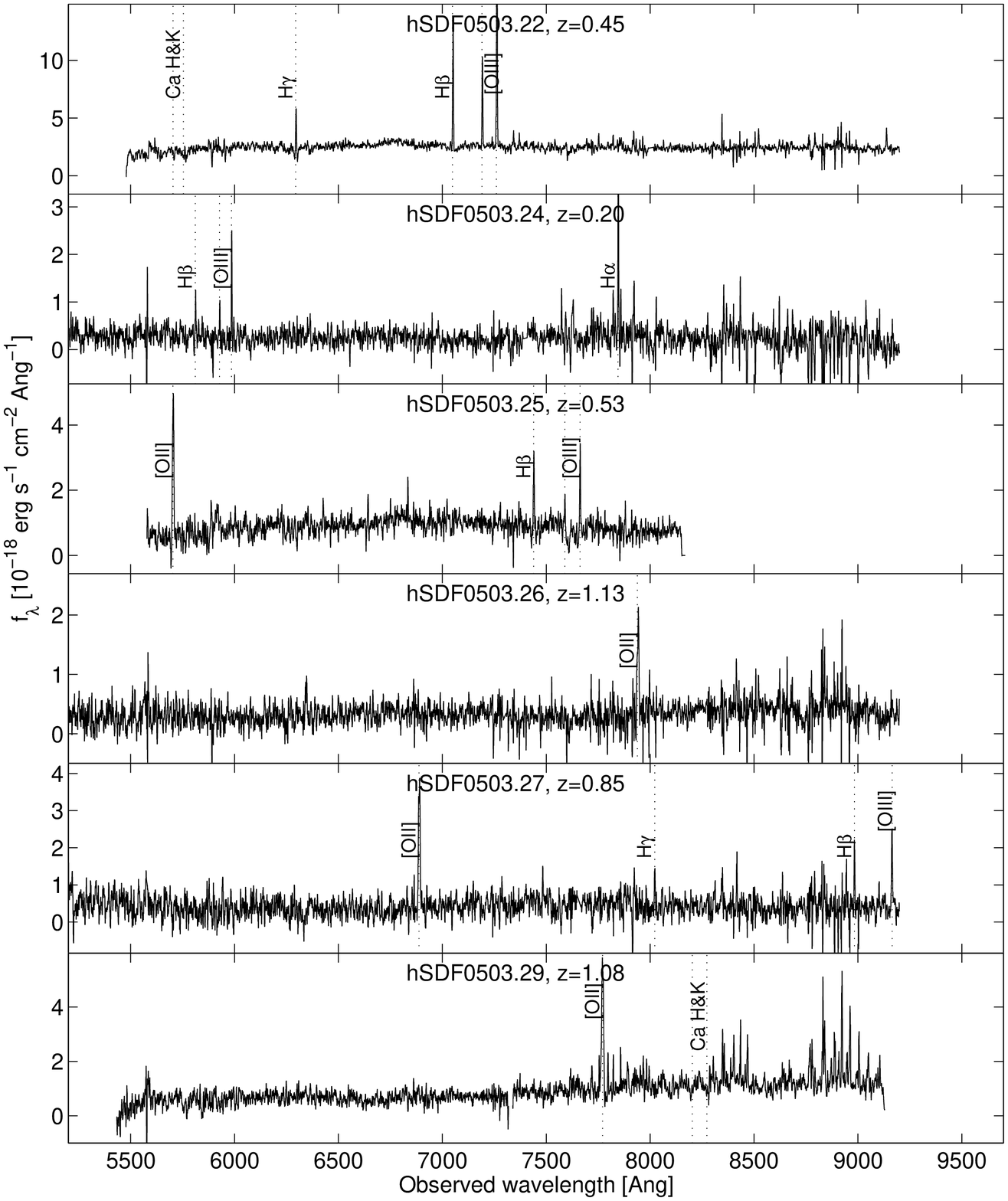}
\end{figure*}
\begin{figure*}
\center
\includegraphics[width=0.9\textwidth]{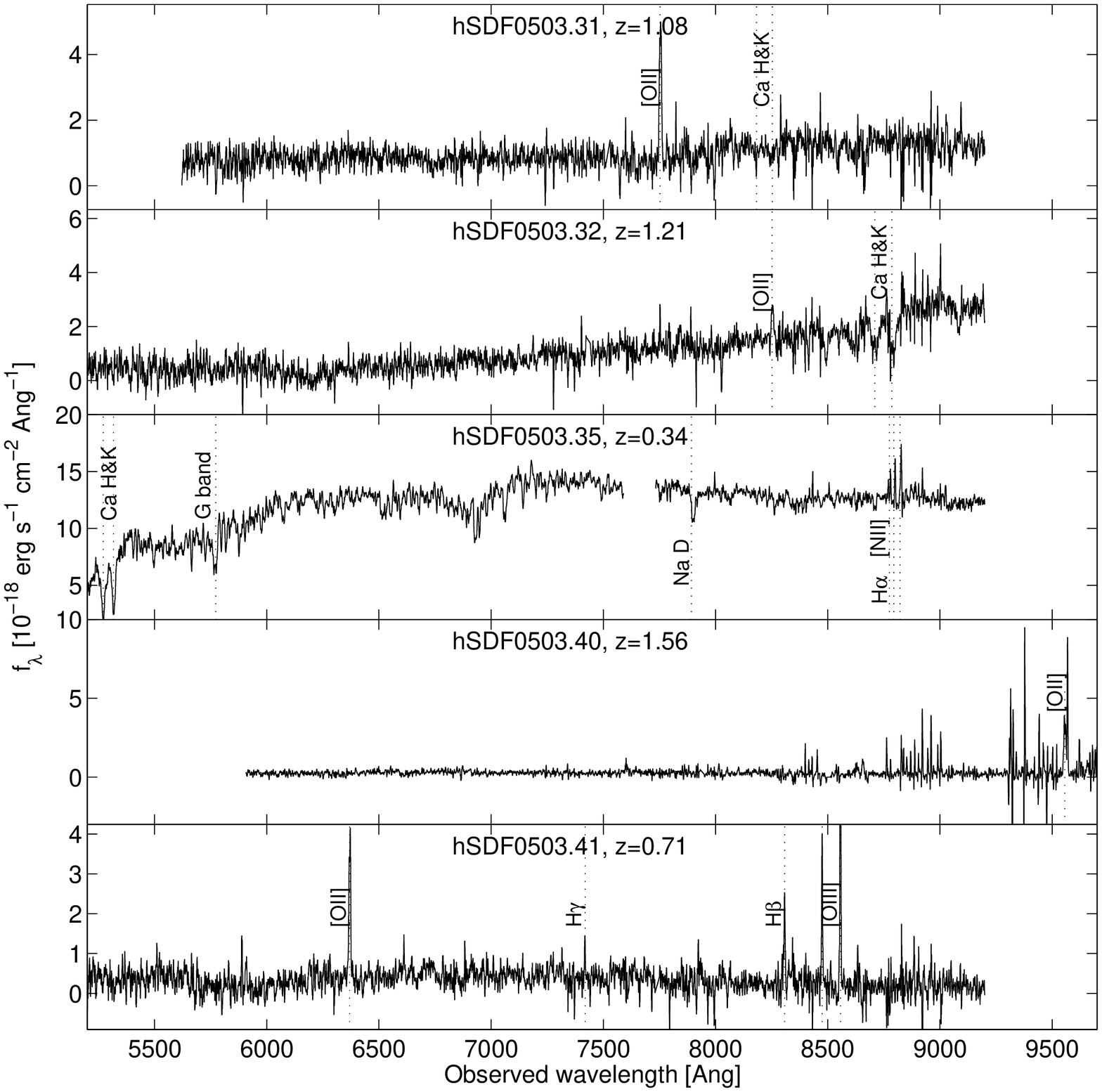}
\caption{Keck DEIMOS spectra for 17 of the SN host galaxies. For most
galaxies, the redshift is securely determined by several emission or
absorption lines. In hSDF$0503.03$, hSDF$0503.26$, and hSDF$0503.40$, the
single line is double peaked, making the identification as
[\mbox{O\,{\sc ii}}] $\lambda\lambda$3726, 3729 quite robust. 
Regardless, any other
assignment predicts the presence of other lines which we do not detect.
Gaps in the spectra correspond to regions with strong night-sky line
subtraction residuals.\label{f:host_spec}}
\end{figure*}

We find the redshifts of the host galaxies using a combination of
photometry and spectroscopy.  Out of 19 host galaxies with
$R<25~\mathrm{mag}$, the approximate practical limit for spectroscopy,
we observed 18, and managed to measure 16 redshifts.  We also measured
redshifts for a fainter host, hSDF$0503.03$, with $R=25.8$ mag. The
spectra are shown in Figure \ref{f:host_spec}.  The spectra of
hSDF$0503.16$ and hSDF$0503.17$, which were presumably bright enough
for Keck ($R=23.56$~mag and $R=24.91$~mag, respectively), did not yield
a conclusive redshift, probably due to slit imperfections that impair
proper sky-line subtraction. We note that no spectrum shows an AGN
signature (except hSDF$0503.35$ that hosts a LINER offset from the SN,
see \S\ref{class}).

The highest spectroscopic redshift we obtained is $z=1.564$, for
hSDF$0503.40$, slightly higher than the redshift of the host of
SN 2003ak found by GOODS at $z=1.551$ \nocite{RIESS_04}({Riess} {et~al.} 2004a). This makes
SNSDF$0503.40$ the SN with the highest host spectroscopic redshift
reported to date. SN 1997ff at $z \approx 1.7$, like other SNe
in our sample having comparable photometric redshifts, 
does not have a spectroscopic SN or host redshift
\nocite{riess_01}({Riess} {et~al.} 2001).

In order to assemble a training set for the derivation of photo-$z$,
we measured 145 spectroscopic redshifts (123 of which are secure) for
galaxies chosen randomly near the positions of the SN hosts, at first
with no selection criteria other than sufficient brightness for
spectroscopy ($i'<25~\rm{mag}$).  In the last Keck run, we
preferentially selected galaxies with photometric redshifts between
1.5 and 2, in order to better sample that region of parameter space.
The redshifts for these 123 galaxies are listed in Table
\ref{t:rand_gal} in the online version of this paper.  We supplement
these data with spectroscopic redshifts for a sample of SDF galaxies
that were obtained for other SDF-related projects \nocite{kashikawa_03,kashikawa_06,shimasaku_06}(e.g.,
 {Kashikawa} {et~al.} 2003, 2006; {Shimasaku} {et~al.} 2006), usually selected for
their excess flux in the narrow-band filters $NB816$ and $NB921$
(designed to detect Ly$\alpha$ emission lines at redshifts near 5.7
and 6.6, respectively). From this sample we select 196 galaxies with
secure $z<2$ redshifts.  This results in a set of 319 galaxies with
secure redshifts that allows us to tune and test the photo-$z$
determination. While this sample is quite extensive, it is probably
biased toward galaxies with bright emission lines, for which
spectroscopic redshifts are easier to measure.

\begin{table*}

\scriptsize
\begin{minipage}{\textwidth}
\center
\begin{tabular}{l|c|c|c|c|c}
\hline
\hline
{ID}  & {$\alpha$ (J2000)} & {$\delta$ (J2000)} &
{$R$ mag} & {Redshift} & {Instrument$^a$}\\
\hline
   1 & 13:23:51.556 & +27:11:51.53 & 24.3 & 1.488 & 4 \\
   2 & 13:23:40.027 & +27:13:01.35 & 23.7 & 0.287 & 4 \\
   3 & 13:24:37.571 & +27:13:24.44 & 23.4 & 0.736 & 2 \\
   4 & 13:24:21.800 & +27:13:22.62 & 23.1 & 0.530 & 2 \\
   5 & 13:24:30.644 & +27:13:37.35 & 25.1 & 1.466 & 2 \\
\nodata\\
\nodata\\
\hline
\end{tabular}
\end{minipage}
\caption{Keck Spectroscopy of SDF Galaxies. The full table is
available in the electronic version of the paper.
$^a$
Instrument and observation date: 1 = LRIS, 12 Jan. 2007; 2 = DEIMOS,
21 Jan. 2007; 3 = DEIMOS, 16 Mar. 2007; 4 = DEIMOS, 12 Apr. 2007.}
\label{t:rand_gal}
\end{table*}

For the derivation of the SN host photometric redshifts, we have used the
recent version of the Zurich Extragalactic Bayesian Redshift Estimator
\nocite{Feldmann_06}(ZEBRA;  {Feldmann} {et~al.} 2006).  Briefly, 
ZEBRA works as follows.
First, the photometry catalog is corrected for systematic errors by
finding residuals in a first fit to a set of basic galaxy spectral
templates. Next, the templates are corrected using a training set with
spectroscopic redshifts. Finally, the redshifts are derived using a
Bayesian methodology, where the redshift and best-fitting template for
each galaxy are found iteratively, while using the resulting
distributions (both in redshift and template space) as priors. Our
initial templates are the same ones used by \nocite{Benitez_00}{Ben{\'{\i}}tez} (2000). For
the correction of the photometry, we measured all the objects in the
SDF, following the procedure described in \S\ref{sample} for the
photometry of the SN host galaxies.  The six basic templates that are
used in BPZ (for galaxy types elliptical, Sbc, Scd, irregular, and two
types of starburst) were interpolated to produce three intermediate
galaxy types between every two consecutive templates. The resulting 21
templates were then corrected using a partial (early) version of the
training set described above, of about 200 galaxies. The correction
was done in three redshift bins ($z<0.5$, $0.5<z<0.9$, $z>0.9$).  This
resulted in 84 templates (the 21 original interpolated ones plus 21
for each redshift bin).  The resulting posterior redshift distribution
of all the galaxies in the SDF was used as a prior for the SN host
redshifts. We set the limit on the absolute rest-frame $B$-band
magnitude of the galaxy to lie, conservatively, between $-26$ and
$-10$.

\begin{figure}
\center
\includegraphics[width=0.5\textwidth]{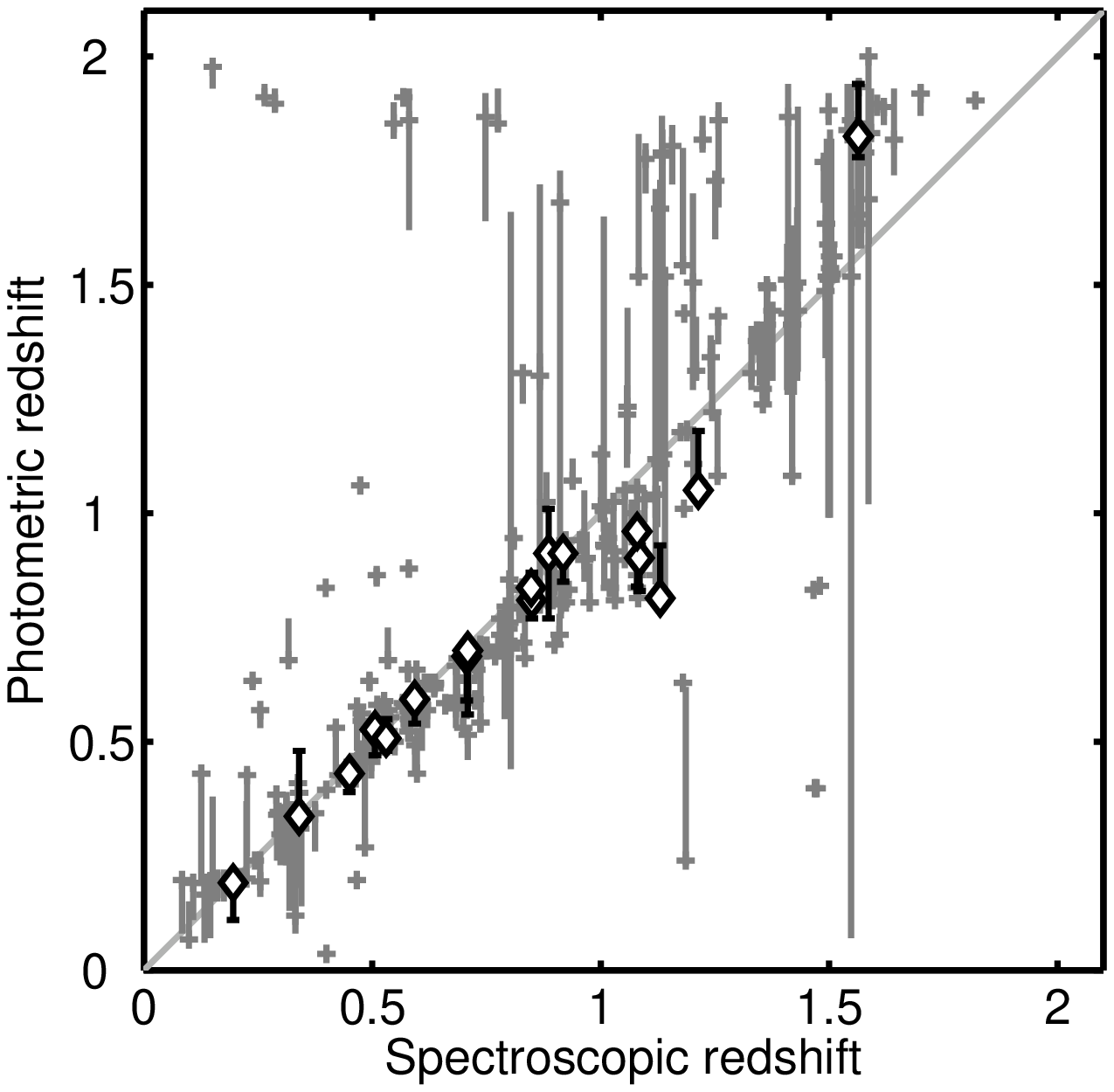}
\caption{Comparison of the spectroscopic redshifts and the ZEBRA
photometric redshifts for the random sample of 319 galaxies (grey
crosses), and for 17 SN host galaxies (empty diamonds).
Error bars indicate the 68.3\% confidence interval of the $z$-pdf for each galaxy.
The scatter of the random sample with photo-$z$ smaller than $1.8$ is
$\sigma_z/(1+z)=0.08$ after excluding five $4\sigma$ outliers.  For the SN host
sample the scatter is $\sigma_z/(1+z)=0.06$.\label{f:phot-spec}}
\end{figure}

Figure \ref{f:phot-spec} shows the results of using ZEBRA on our
training set. First, one can see that the photo-$z$ determination is
usually in good agreement with the spectroscopic redshift.  On the
other hand, photo-$z$ values above 1.8 are many times unreliable for
this training-set sample.  Restricting ourselves to photometric
redshifts smaller than $1.8$, where 93\% of our sample is located (and
probably all of our SNe), leaves us with 296 galaxies, with a
dispersion of $\sigma_z/(1+z)=0.08$ after rejecting five $4\sigma$
outliers (less than 2\% of this sample). The scatter for
our sample of 17 SN host galaxies is somewhat smaller, $\sigma_z/(1+z)=0.06$,
without a single outlier rejected. Application of the
Kolmogorov--Smirnov test to the two samples of residuals indicates that
the difference is  not significant.

Improvement to the photo-$z$ derivation could
come from an extension of the imaging data to the near-infrared, as
the rest-frame 4000--8000~\AA~range is shifted, at $z\approx1.8$, into
the $J$, $H$, and $K$ bands. With existing wide-field
near-IR imagers \nocite{casali_01}(e.g., WFCAM on the UKIRT,  {Casali} {et~al.} 2001), and
certainly future ones \nocite{autry_03}(e.g., NEWFIRM on the NOAO
telescopes,  {Autry} {et~al.} 2003), deep imaging in those bands of well-studied
fields, such as the SDF, would be extremely useful for projects
such as ours.

\begin{figure}
\center
\includegraphics[width=0.5\textwidth]{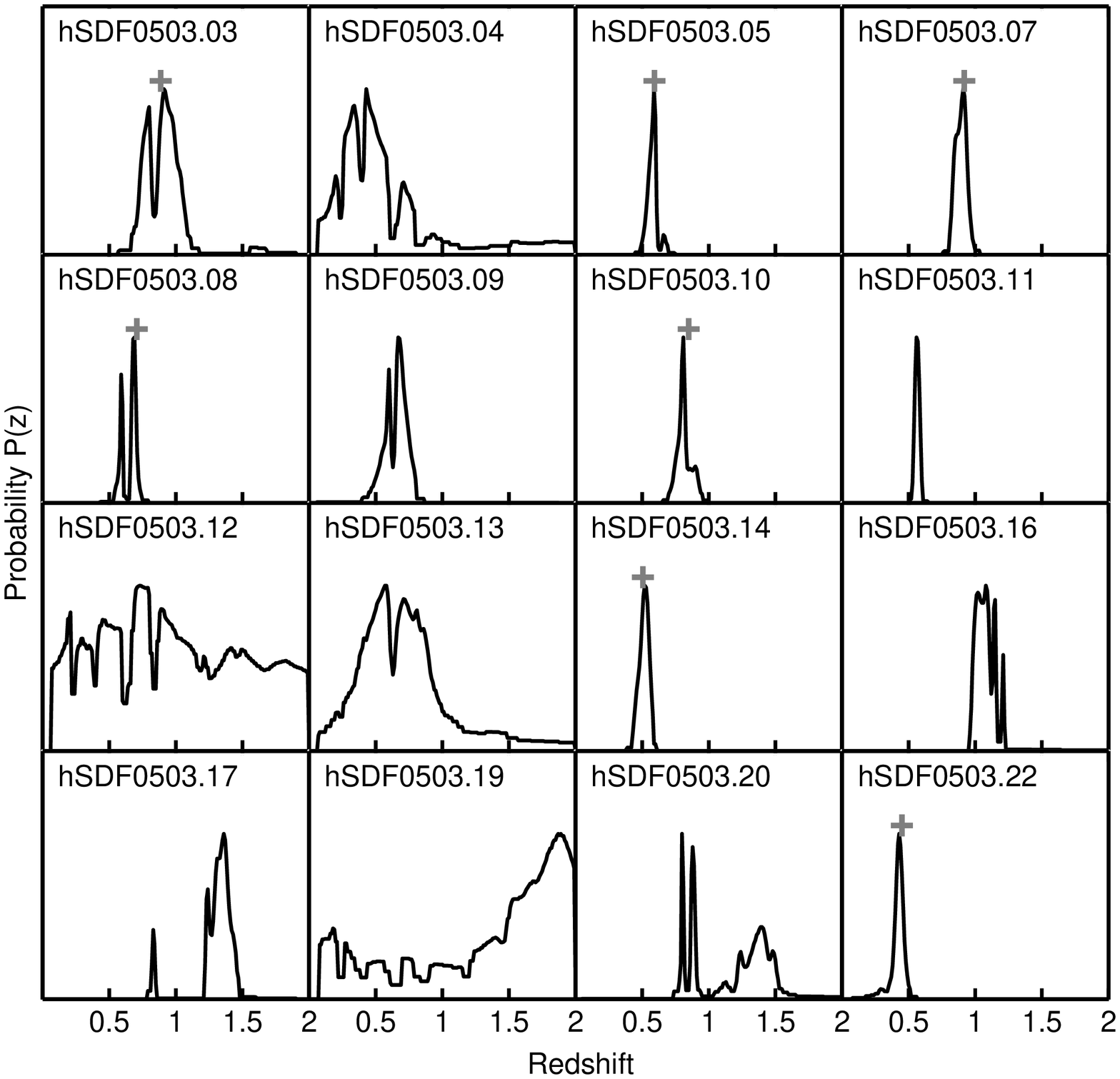}
\includegraphics[width=0.5\textwidth]{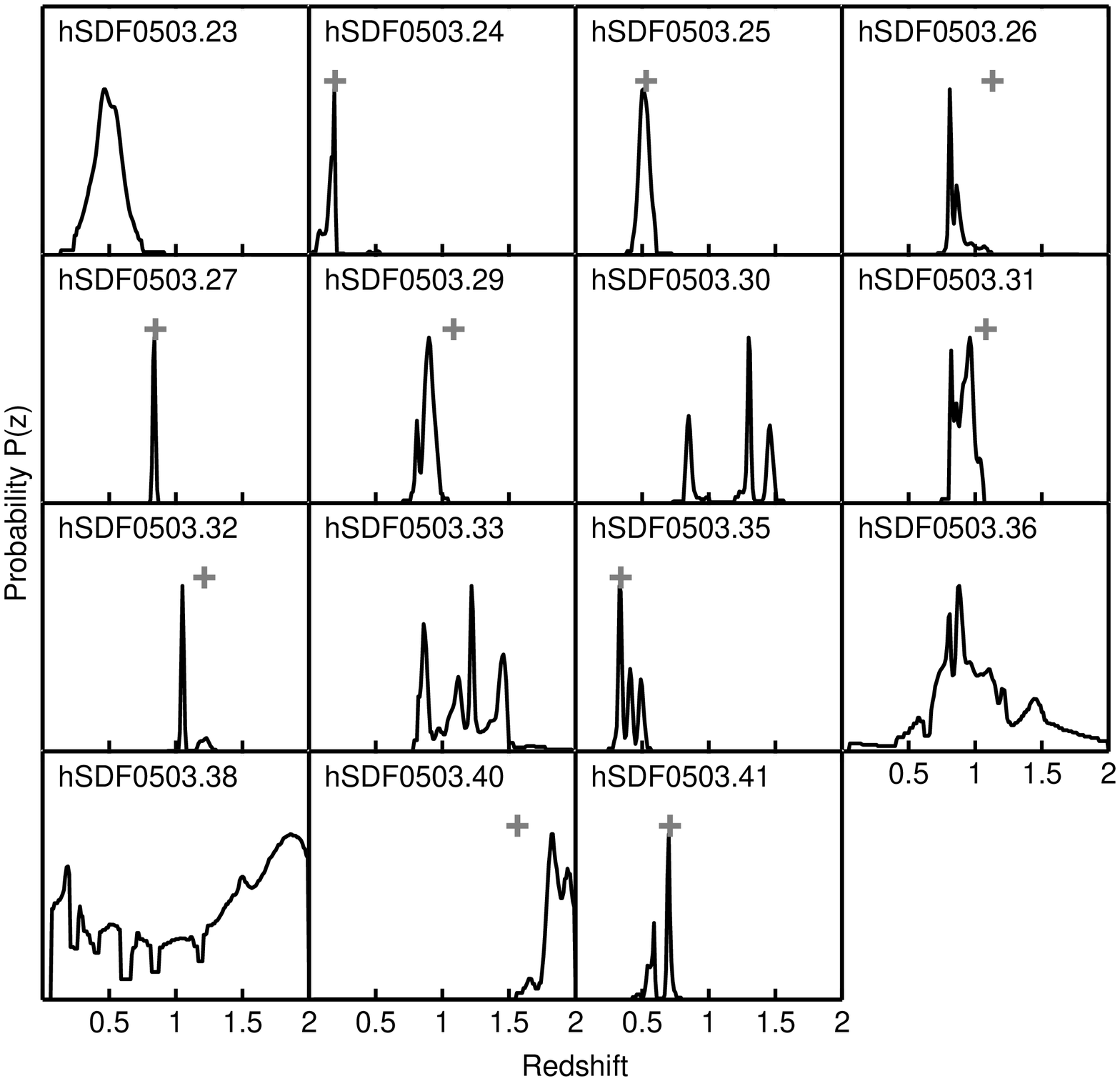}
\caption{Redshift probability density functions ($z$-pdfs) of the SN
host galaxies, as calculated with ZEBRA, and used as priors for the SN
classification. Grey crosses mark the spectroscopic redshifts, when
available. Notice the complex shapes of the $z$-pdfs, and the
agreement in most cases between the most probable photo-$z$ and the
spectroscopic redshift.  All $z$-pdfs have a total probability of 1;
the vertical scale varies from frame to frame.\label{f:pdfs1}}
\end{figure}

The output from ZEBRA that we use is the redshift probability
distribution function ($z$-pdf) for each host galaxy, which is obtained
by marginalizing the full posterior distribution over all
templates. The resulting $z$-pdfs of our host sample are shown in Figure
\ref{f:pdfs1}.  The $z$-pdf can have a complex structure --- it is
not necessarily a single peaked, well-behaved, function. For example,
the distribution for hSDF$0503.03$ is clearly bimodal, while that of
hSDF$0503.12$ contains very little information, with all redshifts
almost equally likely. Using the full $z$-pdfs in the subsequent
analysis allows us to take this uncertainty into account.  Out of the
31 host $z$-pdfs, 23 have widths $w_z/(1+z)<0.08$, where $w_z$ is the
standard deviation of the best-fitting Gaussian to the $z$-pdf,
while only three have $w_z/(1+z)>0.2$ (namely hSDF$0503.12$,
hSDF$0503.19$, and hSDF$0503.38$).  For 16 of the 17 host galaxies for
which we have a spectral redshift, it is very close (usually nearly
identical) to the photo-$z$, with $\Delta z/(1+z)\leq0.1$, while for
the remaining galaxy (hSDF$0503.26$) it differs by only $\Delta
z/(1+z)=0.15$.  For these 17 galaxies, we use in the subsequent
analysis a Gaussian $z$-pdf centered on the measured spectroscopic
redshift, with a width $w_z=0.01$. For the two hostless SNe
(SNSDF$0503.15$ and SNSDF$0503.18$), we use a flat redshift
prior between 0 and 2.
Table \ref{t:hosts} lists the
measured properties of the SN host galaxies, including the
best-fitting redshifts and template types.

\begin{figure}
\center
\includegraphics[width=.45\textwidth]{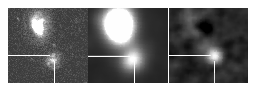}
\caption{Left: image section of $HST$-ACS image of the host galaxy of
SNSDF$0503.31$.  Centre: $R$-band SuprimeCam image prior to SN
explosion.  Right: difference between SuprimeCam images. There is a
clear offset between the SN and its host centre, a face-on spiral
galaxy.\label{f:cand_31hst}}
\end{figure}

The region of the SN host galaxy hSDF$0503.31$ was imaged with {\it
HST}/ACS. As seen in Figure \ref{f:cand_31hst}, this host is a face-on
spiral galaxy, consistent with the photo-$z$ best-fitting template,
with the SN clearly offset from the centre by $0\farcs35$.  While this
$z=1.08$ galaxy seems to be associated with the brighter spiral seen
in the figure, the photo-$z$ of the bright galaxy is much smaller,
$z \approx 0.5$.

\section{Supernova Classification}\label{class}

The next step in our analysis is the classification of the SNe into
types, based solely on our single-epoch photometry and the redshift
information for the hosts.  P07 have recently developed for this
purpose an algorithm, the SN Automatic Bayesian
Classifier (SN-ABC).  Briefly, the SN-ABC is a template-fitting
routine that compares the SN magnitudes to synthetic magnitudes
derived from a library of template SN spectra of different types,
ages, redshifts, and extinctions. The input to SN-ABC is the
photometry of the SN and the $z$-pdf of its host, which is
used as a redshift prior. The output is the
probability that the SN is of Type Ia, $P_{Ia}$, compared to the
probability it is a core-collapse event, $P_{CC}=1-P_{Ia}$, a
posterior $z$-pdf, and a $\chi^2$ value. The  $\chi^2$ criterion
permits rejecting
candidates with colours unlike those of SNe, usually peculiar SNe,
AGNs, or other transients or variables, that were not rejected in
previous steps based on their variability characteristics.
As shown in P07, more than half of the AGNs can be rejected in this manner.
A SN with
value of $P_{Ia}>0.5$ is considered to be of Type Ia, while objects
with $P_{Ia}<0.5$ are classified as CC-SNe. As shown in P07, $P_{Ia}$
also serves as a confidence estimator. The closer it is to unity
(zero), the more secure is the classification as a SN~Ia (CC-SN).

P07 tested the algorithm on different real and simulated datasets. The
SN-ABC successfully classifies 97\% of the SNe~Ia from the Supernova
Legacy Survey \nocite{ASTIER_SNLS06}(SNLS;  {Astier} {et~al.} 2006), and 85\% of the Type
II-P SNe. Similar success fractions were obtained for the GOODS SNe,
and for a simulated sample.  These simulations also show that for Type
Ib/c SNe, success fractions are smaller (about 75\%), while Type IIn
SNe are often misclassified as SNe Ia. The resulting purity of a SN
sample depends on the details of the survey. Consequently, in
\S\ref{contam} we perform simulations with parameters tailored to our
survey, and use them to ``debias'' our sample, and to determine the range
of type distributions consistent with the data.

Table \ref{t:SNe} lists the SNe we have found, their redshifts, and
their ABC classifications.
Among the 33 SNe, 22 are classified as
SNe~Ia by the algorithm.  Among these 22 SN~Ia classifications, 16
have $P_{Ia}$ values higher than 0.7, while 6 of the 11 CC-SNe are
classified with similar confidence, $P_{CC}>0.7$.

Two SNe, SNSDF$0503.24$ and SNSDF$0503.35$, have extreme $\chi^2$ values
of 22.2 and 30.1, respectively. As discussed above, this could be indicative of a non-SN object, since,
as shown in P07, the SN-ABC assigns  high  $\chi^2$ values to most AGNs.
However, both are well offset from the centres of their host galaxies.
The host hSDF$0503.35$ is a ``grand-design'' face-on spiral galaxy (Fig. 
\ref{f:stamps}) having a LINER spectrum \nocite{ho_97}({Ho}, {Filippenko} \& {Sargent} 1997) with an H$\alpha$ 
luminosity of $\sim10^{39}$~erg$\,$s$^{-1}$, typical of such objects.
Thus, these are probably not AGNs (see, however, \nocite{AGY_99dd}{Gal-Yam} 2005 and
Gal-Yam et al. 2007, in preparation,
who found a background AGN projected near a lower-redshift galaxy).
Both of their respective hosts
have relatively low (spectroscopic) redshifts, and are, in fact, the 
lowest-redshift SNe in our sample.
Both are classified as CC-SNe, with high confidence, which in the context of the SN-ABC means that
their colours resemble significantly more those of a SN~II-P than a SN~Ia.
However, the large $\chi^2$ values imply that neither SN type is a good fit.
Since these are unlikely to be SNe~Ia, we keep them in our CC-SN sample.
Nevertheless, we will continue to monitor these positions in future observations,
in order to test the background AGN option.

We also note that hSDF$0503.35$ is classified by ZEBRA as an elliptical
galaxy, in contradiction with its
spiral morphology. This is probably due to the photometry being
dominated by the bulge,
while the SN could be in the disc. Another CC-SN host, hSDF$0503.23$, 
is also classified as an elliptical,
suggesting that, here as well,
the classification, either of the host galaxy, or of its SN, is erroneous.
However, we expect a fraction of the SNe to be misclassified, and we deal
explicitly with the resulting uncertainties and biases in \S\ref{contam}.

The posterior redshift, derived from the combined fit of each SN and
its host, is generally quite close to the prior (host-galaxy-only
based) redshift, but as shown in P07, these posterior redshifts can
sometimes be biased, and therefore we generally do not use them in the
next step.  A clear example of disagreement between a prior and a
posterior redshift is SNSDF$0503.38$, which has a prior redshift of
1.87, while the SN photometry combined with this prior yields a
classification as a CC-SN at $z=0.3$. This apparent discrepancy is
easily understood if one notices that the $z$-pdf for the host galaxy
of this SN is very wide and unconstraining, and hence the best-fitting
redshift for the host is quite meaningless.  For this object, as well
as for SNSDF$0503.12$ and SNSDF$0503.19$ that have similarly wide
$z$-pdfs, and the hostless SNSDF$0503.15$ and SNSDF$0503.18$,
we use the posterior redshifts that take into account the
SN properties.

For a rough idea of the properties of the SNe, we use the peaks of the $z$-pdfs
as best-fitting redshifts, and
find approximate rest-frame absolute magnitudes
of the SNe. First, we fit the de-redshifted fluxes with a linear function, to
describe the spectral energy distribution (SED) of the SN. 
We then perform synthetic photometry in the $B$ band,
unless rest-frame $B$ does not fall within the observed $R$, $i'$, or $z'$ bands.
In such a case, we derive rest-frame $U$ or $V$ absolute magnitudes,
at high or low redshift, respectively. The resulting absolute magnitudes,
listed in Table \ref{t:SNe}, range
between $-13.2$ and $-17.3$ for the CC-SNe, and between $-15.1$ and $-19.5$ for the SNe~Ia.

We divide the samples of SNe~Ia and CC-SNe into four redshift bins
($z<0.5$, $0.5<z<1.0$, $1.0<z<1.5$, and $1.5<z<2.0$), 
using the peak of the $z$-pdfs as best-fitting redshifts.
These ``raw'' distributions are shown in Fig. \ref{f:zdist} and 
listed in Table \ref{t:dist}.

\section{Uncertainties, Sample Contamination, and Derivation of Intrinsic Supernova Type and Redshift
Distributions}\label{contam}

The ABC classification can introduce biases due to the dependence of
the success fraction on the intrinsic parameters (type, age, redshift)
of the SNe. We therefore use Monte Carlo simulations to determine the
allowed range, and the most likely ``true" redshift and type
distribution, of the SNe (as we would have observed if we had perfect
spectroscopic data for all of the SNe and their hosts).

Following P07, we have randomly generated $10,000$ SN spectra of each
of the main SN types: Ia, Ibc, II-P, and IIn \nocite{FILIPPENKO_97}(see, e.g.,  {Filippenko} 1997).
We calculate the synthetic magnitudes of the fake SNe using the
SEDs from the spectral templates of
\nocite{NUGENT_02}{Nugent}, {Kim} \& {Perlmutter} (2002) for SNe~Ia, Ib/c, and II-P.  Since the
\nocite{NUGENT_02}{Nugent} {et~al.} (2002) spectra of SNe~IIn are theoretical blackbody
SEDs, for this type only we use the templates from \nocite{POZ_TP1}{Poznanski} {et~al.} (2002).
Absolute magnitudes and their dispersions are taken from
\nocite{DAHLEN_SNR04}{Dahlen} {et~al.} (2004). The scatter in colour within each type is
applied to the CC-SNe by adding an intrinsic, normally distributed,
noise with a standard deviation of $\sigma=0.2~\mathrm{mag}$, the
value used by \nocite{SULL_TP}{Sullivan} {et~al.} (2006a).  The SNe~Ia we simulate are assigned
different stretch values $s$, following the method described by
\nocite{SULL_TP}{Sullivan} {et~al.} (2006a). We simulate a Gaussian distribution of stretch values with
an average of $s=1$ and a dispersion of $\sigma=0.25$, truncated
outside the range $0.6 \leq s \leq 1.4$. We model the
stretch-luminosity relation using the formalism
$M_{Bc}=M_B-\alpha(s-1)$ \nocite{Perlmutter_99}({Perlmutter} {et~al.} 1999), where $M_{Bc}$ and
$M_{B}$ are the corrected and uncorrected $B$-band absolute
magnitudes, respectively, and the correlation factor is
$\alpha=1.47$. We also apply colour-stretch corrections using the
method presented by \nocite{Knop_03}{Knop} {et~al.} (2003), by dividing the template spectra
of normal, $s=1$, SNe~Ia, by smooth spline functions, in order to
match their rest-frame $UBVRI$ colours to those of SNe with various
stretch values.

To every simulated SN we assign an epoch, a host-galaxy extinction,
and a redshift. The epoch is drawn from a uniform distribution, while
the host extinction, $A_V$, is drawn from the positive side of a
Gaussian distribution with maximum at zero, $\sigma=0.2$ mag for SNe~Ia,
and $\sigma=0.5$ mag for CC-SNe, truncated at $A_V=1$ mag. The redshifts
are drawn from the general galaxy population in the SDF using the
following scheme.  A realistic photometric redshift is modeled, by
drawing, for half of the SNe, the $z$-pdf of a random SDF galaxy with
$z<2$. To the other half of the simulated SNe we assign a
spectroscopic redshift, by using a $\sigma=0.01$ Gaussian
$z$-pdf. This mimics the redshift determination characteristics of the
real sample.  We have measured the widths of the simulated sample's
$z$-pdfs, as well as of the real host-galaxy sample, by fitting them
with Gaussians, and find that the simulated and real samples
have similar distributions of widths.  We draw a specific redshift for
each simulated SN from its $z$-pdf, and use it to redshift the
simulated spectrum\footnote{This is the most straightforward way to
model reality, where the order is reversed, i.e., an object is at
a particular redshift, and one measures its $z$-pdf.}.

We then fold the simulated SN spectra through the observational setup
of our SDF observations, with their three bandpasses, photometric
errors, and limiting magnitudes.  We have measured the mean
photometric errors of the SDF sample in each band as a function of
magnitude, and assuming that the noise is normally distributed, have added
it to each object. As with the real sample, we keep only objects with
$z'$-band magnitudes brighter than $26.3$, leaving $4,000$ of each
type.  The simulated SNe are then blindly classified by the ABC into
SNe~Ia or CC-SNe, producing a mapping of success fraction vs. redshift
for each SN type.  We find the success fractions of the ABC in four
redshift bins ($z<0.5$, $0.5<z<1.0$, $1.0<z<1.5$, and $1.5<z<2.0$) for
each of the four SN types.

\begin{figure}
\center
\includegraphics[width=0.5\textwidth]{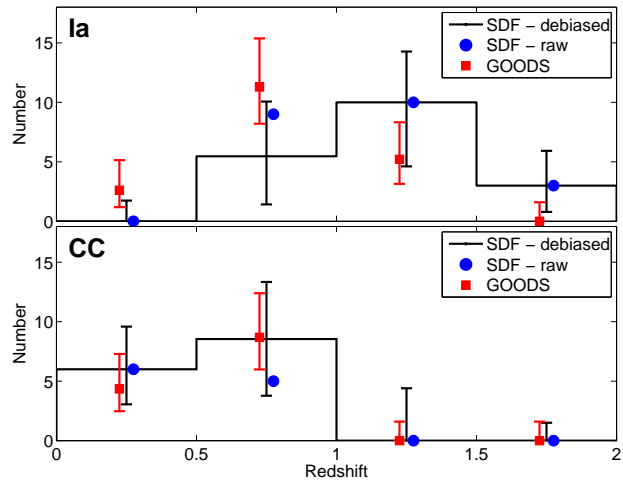}
\caption{Observed, raw (circles; slightly offset to the right for
clarity), and debiased (lines), redshift distributions of SDF SNe~Ia
and CC-SNe, compared to the GOODS sample (squares; offset to the
left). SDF error bars mark $1\sigma$ classification and Poisson
errors, added in quadrature. GOODS error bars denote $1\sigma$
Poisson uncertainties.  The debiased SDF and GOODS distributions are
consistent to within the uncertainties, but there is a suggestion
of more high-$z$ SNe~Ia in the SDF.\label{f:zdist}}
\end{figure}

The success fractions are applied to the set of all possible intrinsic
type distributions (e.g., 30\% SN~Ia, 30\% SN~II-P, 20\% SN~Ib/c, and
10\% SN~IIn).  Using steps of 2.5\%, there are $12,341$
possibilities to distribute 100\% of the observed SNe, in each
redshift bin, among the four types.  We compute, for each of these
possibilities, the resulting fraction of objects that are labelled as
SNe Ia in the sample (i.e., the sum of the fraction of SNe~Ia that
were correctly classified), and of the fraction of CC-SNe that were
misclassified as SNe~Ia, as obtained by the average success fractions
for each type of SN, in each redshift bin.  Next, for each scenario,
in each redshift bin, we find the binomial probability to draw the
observed raw number of SNe~Ia out of the total number of SNe in the bin,
given the expected SN~Ia fraction for that scenario.  Since, in this
scheme, the SN~Ia fractions do not have equal prior probability (e.g.,
there is only one model with 100\% SNe Ia, while there are many with
50\% SNe~Ia and various fractions for the CC-SN subtypes), we multiply
each probability by a weight function that is inversely proportional
to the number of scenarios with the same SN~Ia fraction. Finally, we
marginalize over all the scenarios in each redshift bin, and obtain
the probability density function for the true SN~Ia fraction in that
bin. From the probability distribution function of the SN~Ia
fraction in each bin, we extract the most probable value, at the peak
of the distribution, and the $1\sigma$ uncertainties defined as the region
that includes $68.3\%$ of the probability density. We add
to this error the corresponding fraction of the $1\sigma$
Poisson uncertainty in that bin --- that is, the Poisson uncertainty for the
total number of SNe in the bin multiplied by the relevant fraction of
SNe~Ia.  The most probable SN~Ia fraction comes out the same as the 
SN~Ia fraction given by the SN-ABC in two out of the four redshift
bins. Only in the second redshift bin, $0.5<z<1.0$, is the debiased
most-probable fraction of SNe~Ia reduced from 0.64 to 0.39.  The raw
and debiased redshift distributions of the SNe~Ia and CC-SNe are
listed in Table \ref{t:dist}, and shown in Figure \ref{f:zdist}. We
discuss the results in the following section.

\begin{table*}
\begin{minipage}{\textwidth}
\center
\begin{tabular}{l|l|l|l|l}
\hline
\hline
{Sub-sample} & {$z<0.5$} & {$0.5<z<1.0$} & {$1.0<z<1.5$} & {$1.5<z<2.0$}\\
\hline
SN Ia - (raw)      & 0 & 9 & 10 & 3 \\
SN Ia - de-biased$^a$           &$0.0^{+1.7+0.0}_{-0.0 -0.0}$ & $5.5^{+4.2+1.9}_{-3.8 -1.4}$ &
 $10.0^{+0.0 +4.3}_{-4.4 -3.1}$ & $3.0^{+0.0 +2.9}_{-1.5 -1.6}$\\
CC-SN - (raw)      & 6 & 5 & 0 & 0 \\
CC-SN - de-biased$^a$           &$6.0^{+0.0 +3.6}_{-1.7 -2.4}$ & $8.5^{+3.8 +2.9}_{-4.2 -2.3}$ & $0.0^{+4.4 + 0.0 }_{-0.0 -0.0}$ & $0.0^{+1.5 +0.0}_{-0.0 -0.0 }$ \\
Total   & 6 & 14 & 10 & 3\\
\hline
SN Ia rate$^b$ [10$^{-5}$\,Mpc$^{-3}\,$yr$^{-1}$] & $0.0^{+2.4}_{-0.0}$ &$4.3^{+3.6}_{-3.2}$ &$10.5^{+4.5}_{-5.6}$ &$8.1^{+7.9}_{-6.0}$ \\
\hline

\end{tabular}
\end{minipage}
\caption{Supernova Redshift Distribution.
$^a$Errors are $1\sigma$ classification and Poisson uncertainties,
respectively. $^b$ SN~Ia rate errors are classification and Poisson uncertainties added in quadrature}
\label{t:dist}
\end{table*}

\section{Comparison to the GOODS Supernova Sample}\label{s:goods_comp}

Before calculating rates, which requires some assumptions and modeling
of the SN properties,
we compare our SN sample to the most similar one
in terms of depth and numbers, the GOODS SN sample. As part of the
GOODS project, two fields of about $150~ \textrm{arcmin}^2$ each were
observed with the ACS camera on {\it HST} every 45 days, over five
epochs, in order to search for SNe. We have compiled from
\nocite{RIESS_04}{Riess} {et~al.} (2004a) and \nocite{STROLGER_04}{Strolger} {et~al.} (2004) all available photometry for
the 42 SNe found in the GOODS fields. These SNe cover a redshift range
of $0.2$ to $1.55$, with a median of $0.76$. The SNe have partial
light-curve coverage, and not all have spectroscopic redshifts or
types. However, GOODS is a complete sample, without any subsequent
selection.

Our survey and GOODS have different effective areas and observing
strategies, for which we need to account prior to comparison. In terms 
of bands, the surveys are similar.  The filter-plus-system bandpasses
of the Suprime-Cam $z'$ filter and of the {\it HST} F850LP filter
(used by GOODS) have effective wavelengths at $9067\,$\AA~and
$9082\,$\AA, respectively.  The GOODS detection efficiency function,
as described by \nocite{STROLGER_04}{Strolger} {et~al.} (2004), is slightly deeper than our own,
by about 0.1--0.2 mag. We consider only SNe brighter than our 50\%
efficiency limit, $z'=26.3$~mag.  For $z'$-band magnitudes brighter
than 26.3, both surveys have similar efficiencies, and no correction
is needed for a comparison to be made, if we exclude the fainter GOODS
SNe.  The SDF covers $\sim 900~\mathrm{arcmin}^2$, and has a single
epoch, since we searched for SNe only in the March 2005 images. GOODS
covered $300~\mathrm{arcmin}^2$, imaged at five epochs, each separated
by about 45 days, i.e., 2--4 weeks in the typical SN rest frames. This
means that the epochs are not independent and cannot be simply summed,
since most SNe are discovered early and are detected in more than one
epoch. From the photometry in \nocite{RIESS_04}{Riess} {et~al.} (2004a) and
\nocite{STROLGER_04}{Strolger} {et~al.} (2004), we find that, out of 42 SNe, 38 are detected
(i.e., brighter than 26.3 mag, our SDF cutoff magnitude) in $1.45$
epochs, on average, while the remaining 4 are too faint.  Thus, the
effective GOODS area is roughly $5\times300~\mathrm{arcmin}^2 /1.45=
1036~\mathrm{arcmin}^2$, about 15\% more than the SDF. Hence, for a
direct comparison of the two samples, the number of GOODS SNe needs to
be trimmed down from 38 to $38/ 1.15 = 33$.  Due to this reduction
factor, we will present non-integer numbers of SNe from GOODS in the
discussion below.

Thus, after the normalization for the different effective areas, there
are (coincidentally) exactly the same number of SNe in our sample and in
the ``SDF-normalized'' GOODS sample (33), 18.9 of which are SNe Ia
(compared to the most probable value of 18.5 in the SDF debiased
sample) and 14.1 CC-SNe (most probable value 14.5 in the SDF).  In
Figure \ref{f:zdist}, we compare the redshift distributions for the
SN~Ia and CC-SN subsamples from the two surveys.  Since one of the
CC-SNe in the GOODS sample (SN 2002fv) has no redshift information,
we exclude it
from this comparison.  As can be seen, the two samples agree well in
total size, in total type fractions, and in the CC-SN redshift
distribution.

However, the redshift distributions may differ for the
SNe~Ia.  Our lowest-redshift SN~Ia is at $z \approx 0.5$, while GOODS
found SNe~Ia down to $z \approx 0.2$. Furthermore, the SNe~Ia in our
sample reach higher redshifts, up to $z \approx 1.6$, rather than
$z \approx 1.4$, in the ``SDF-normalized'' GOODS sample.
While formally the distributions are consistent, our data
could indicate that the results in \nocite{STROLGER_04}{Strolger} {et~al.} (2004) and
\nocite{DAHLEN_SNR04}{Dahlen} {et~al.} (2004) concerning the paucity of SNe~Ia at high $z$, and
the implications for the SN~Ia delay time, may change once larger
samples are obtained. On the other hand, one of our high-$z$ SNe~Ia,
SNSDF$0503.18$, at $z=1.60$, is hostless, and its redshift is based solely on the posterior fit
from the SN-ABC to the SN photometry. If this redshift is in error, the number of SNe~Ia
in the highest-redshift bin would be more similar to that of GOODS.
At $z=0.55$, where the
fraction of SNe~Ia in our sample was most significantly reduced by the
debiasing of our photometrically classified sample, lies the largest
disagreement between competing measures of the SN~Ia
rates. \nocite{BARRIS_SNR06}{Barris} \& {Tonry} (2006), who used photometric methods to classify
SNe, found a rate 4--5 times higher than measured by \nocite{Pain_02}{Pain} {et~al.} (2002)
and \nocite{Neill_06}{Neill} {et~al.} (2006). \nocite{Neill_06}{Neill} {et~al.} (2006) have suggested that the sample
of \nocite{BARRIS_SNR06}{Barris} \& {Tonry} (2006) may have suffered from contamination by
CC-SNe.  Our raw sample of SNe~Ia in that same bin, without debiasing, would
certainly be contaminated by CC-SNe.

\section{The Type Ia Supernova Rate}
The SN redshift distributions that we have derived can be used
to set constraints on the SFH and SN~Ia delay functions.
However, our SN sample is still small, and considering the similarity in
the redshift distributions of the SDF and GOODS CC-SN samples, we postpone
a CC-SN rate derivation to future papers.
Nevertheless, we calculate SN~Ia rates, to examine how the differences
in the distributions noted in \S\ref{s:goods_comp} propagate into the underlying rates.
We calculate the volumetric rate of SNe~Ia, in four bins,
\begin{equation*}
r_{Ia,i}=\frac{N_{Ia,i}}{\int{\eta(z) \, dV}},
\end{equation*}
where $N_{Ia,i}$ is the number of SNe~Ia in bin $i$, and $\eta(z)$ and $dV$ are, respectively,
the survey control time (see below) and the co-moving volume element (in Mpc$^{-3}$),
as a function of redshift $z$.
The integration is done within the limits of each bin.
The control time, $\eta(z)$, is found as follows.
For every redshift $z$, we calculate the model
$z'$-band SN~Ia light curve, $m_{z}(t)$, and sum
over the survey efficiency, $\epsilon(m)$, at these magnitudes:
\begin{equation*}
\eta(z)=\int{\epsilon(m_{z}(t))\,\frac{dm}{dt}\,dt}.
\end{equation*}
The effective redshift of each bin, $<z>_i$, is defined as
\begin{equation*}
<z>_i=\frac{\int{\eta(z)\,z\,dV}}{\int{\eta(z) \,dV}},
\end{equation*}
i.e., a mean redshift weighted by the volume and the control time.

\begin{figure}
\hspace{-0.6cm}
\includegraphics[width=0.55\textwidth]{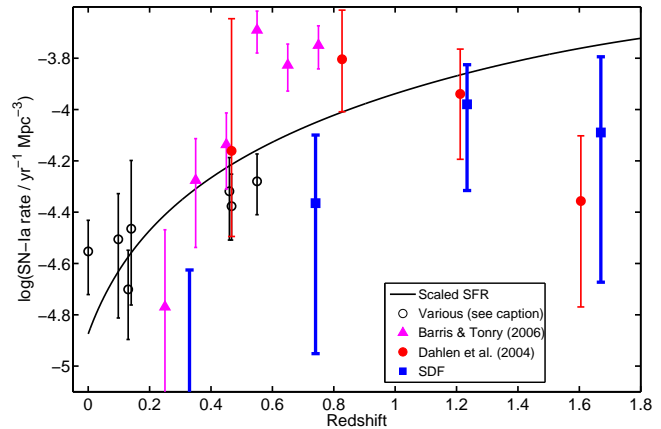}
\caption{Type~Ia SN rate from the SDF, compared to results from the literature
(compiled by Neill et al. 2006).
Empty circles are for Cappellaro et al. (1999), Hardin et al. (2000), 
Pain et al. (2002), Madgwick
et al. (2003), Tonry et al. (2003), Blanc et al. (2004), and
Neill et al. (2006); triangles are for Barris \& Tonry (2006); filled
circles are for Dahlen et al. (2004); and filled squares are for our
derived SDF rate. The line shows the SFH from Hopkins \& Beacom (2006), scaled to fit the $z<0.5$ points.
 \label{f:rate}}
\end{figure}

The uncertainties are the Poisson and classification uncertainties
of the sample, added in quadrature.
To investigate how errors in the efficiency function plotted in Fig. \ref{f:eff}
might propagate to our rate measurement, we have fitted the efficiency curve
with a Fermi-Dirac-like step-function, derived the uncertainties in its parameters,
and have repeated the
rate calculation hundreds of times with randomly drawn efficiency curves.
The resulting $1\sigma$ dispersion in the SN rates is of the order
of $10^{-6}$Mpc$^{-3}$yr$^{-1}$, i.e., about 30 times smaller than the main sources of uncertainty.
If, for some unknown reason, our detection efficiency curve is offset by $0.1$~mag,
the derived rates would change by about 15\%, again less than
the main sources of uncertainty.
A significantly larger shift of the efficiency function,e.g., by 0.3~mag toward bright magnitudes,
would lower the efficiency to near zero at $z'=26.3$~mag, and would be inconsistent
with our having discovered SNe in this magnitude range
(unless, of course, the intrinsic number of SNe rises sharply precisely at this brightness).
We further neglect dust extinction in this rough derivation of the rate.
As found by \nocite{riello_05}{Riello} \& {Patat} (2005) and \nocite{Neill_06}{Neill} {et~al.} (2006), such an omission
causes the rates to be underestimated by $\sim$25\% at most, which is
smaller than the uncertainties already taken into account.
We also do not attempt at this point to incorporate SNe with different stretch values into
the control-time calculation, which should have a negligible (if any) effect \nocite{Neill_06}(e.g.,  {Neill} {et~al.} 2006).
The resulting rates are listed in Table~\ref{t:dist}.
Figure~\ref{f:rate} shows the rates
compared to other published results \nocite{Neill_06}(as compiled by  {Neill} {et~al.} 2006),
with the SFR from \nocite{hopkins_06}{Hopkins} \& {Beacom} (2006) plotted to guide the eye.
The SFR is scaled to fit the low-redshift ($z < 0.5$) measurements, where there are fewer
conflicting estimations.


In our lowest-redshift bin we have found no SNe~Ia. Consequently, the error
bar  marks the upper limit
arising from classification uncertainty only. Assuming the rate is in the range
$10^{-4.8}-10^{-4.2}$\,yr$^{-1}\,$Mpc$^{-3}$,
we expect between 1 and 4 SNe~Ia, given our survey volume and control time,
which is consistent with us finding no SNe in this small-volume bin.
In the $0.5<z<1$ bin, our measurement appears to be inconsistent with the four highest-rate points
of \nocite{BARRIS_SNR06}{Barris} \& {Tonry} (2006), and of \nocite{DAHLEN_SNR04}{Dahlen} {et~al.} (2004).
However,
we note that these rate measurements show a conspicuous enhancement, relative to
most of the other results, at $0.4<z<0.6$.
We further note that the
error bars on the \nocite{BARRIS_SNR06}{Barris} \& {Tonry} (2006)
dataset are Poisson errors alone, and do not include
any systematic uncertainties, such as classification errors that the sample may suffer from,
at least to some extent.
In the high-redshift bins, our rates are in good agreement with the \nocite{DAHLEN_SNR04}{Dahlen} {et~al.} (2004) results.
Nevertheless, we find a $\sim$50\% higher rate of SNe~Ia in the highest redshift bin.
Overall, the central values of our measurements suggest a SN~Ia rate that is fairly constant with redshift
at high redshift, and could be tracking the SFR.
Such a behavior is expected at high redshift if the 
rate then is dominated by a prompt SN~Ia channel with a short
delay time, and contrasts with the results of \nocite{DAHLEN_SNR04}{Dahlen} {et~al.} (2004), whose
data suggest a declining SN~Ia rate beyond $z=0.8$ \nocite{Scannapieco_05}({Scannapieco} \& {Bildsten} 2005).

\section{Conclusions}

Large samples of high-$z$ SNe are required to resolve critical
incompatibilities between recent results regarding the rates of
SNe~Ia, and their delay time from star formation to explosion. Here we
have presented an initial sample of 33 SNe from a new survey based
on deep, single-epoch, re-imaging of the $0.25~\mathrm{deg}^2$ SDF.  We
have demonstrated that, for SN-rate purposes, such surveys can be
carried out using ground-based observations, at a fraction of the cost
of space-based data, and with the potential for large samples.  Using
Keck spectroscopy for more than half of the host galaxies,
we have shown the reliability of photometric redshifts for such data,
incidentally finding the SN with the highest host spectroscopic redshift
reported to date, at $z=1.564$.

The SNe were classified using a novel Bayesian photometric algorithm,
using solely the SN photometry
in three bands, and the host-redshift information, either
photometric or spectroscopic.
The photometric classification of the SNe introduces biases for which
we correct using simulations.  The debiasing procedure we apply to the
raw sample reduces significantly the fraction of SNe Ia in the
redshift bin $0.5<z<1$, which otherwise would have been heavily
contaminated by CC-SNe. As discussed by \nocite{Neill_06}{Neill} {et~al.} (2006), such
contamination may have affected the sample acquired by
\nocite{BARRIS_SNR06}{Barris} \& {Tonry} (2006), and could explain the discrepancy between their
measurement of the SN~Ia rate and the rates measured by
\nocite{Pain_02}{Pain} {et~al.} (2002) and \nocite{Neill_06}{Neill} {et~al.} (2006).

Our resulting sample is comparable to the GOODS SN sample, and is in
good agreement with it, in terms of the total numbers of SNe~Ia and
CC-SNe, and in terms of the redshift distribution of the CC-SN sample.
However, our sample of SNe~Ia shows less of a decline at high
redshifts than the GOODS sample does.  This trend remains when
comparing the SN~Ia rates we derive to those derived by GOODS.
We find at redshift $z\approx0.7$ a rate which is similar to
the values determined by \nocite{Pain_02}{Pain} {et~al.} (2002), \nocite{TONRY_03}{Tonry} {et~al.} (2003), and \nocite{Neill_06}{Neill} {et~al.} (2006)
at $z\approx0.5$, but barely consistent with the results of
\nocite{BARRIS_SNR06}{Barris} \& {Tonry} (2006) and \nocite{DAHLEN_SNR04}{Dahlen} {et~al.} (2004). At $z\approx1.6$, our rate
determination is consistent with the result of \nocite{DAHLEN_SNR04}{Dahlen} {et~al.} (2004),
but about 50\% higher.

These results, if confirmed by larger
samples, may challenge current conclusions, based on the paucity of
SNe~Ia found by GOODS at high redshift, concerning the SN~Ia rate and the delay
time.  Additional epochs on this field, already being obtained, will
enlarge our SN sample to the hundreds, and will test the reality of the
apparent decline in the SN~Ia rate at $z \gtrsim 1$.  Finally, our
approaches to SN search and photometric classification are likely to
become important in the era of synoptic telescopes such as Pan-STARRS
\nocite{Kaiser_05}({Kaiser} {et~al.} 2005) and the Large Synoptic Survey Telescope
\nocite{Stubbs_04}({Stubbs}, {Sweeney} \& {Tyson} 2004), where the sheer numbers of SNe will require new approaches to analysis.

\section*{Acknowledgments}

We thank Nobuo Arimoto for his contribution to this project.  We are also
grateful to R. Feldmann, C. M. Carrollo, and P. Oesch for kindly
providing us with the ZEBRA code, and for assisting us in its use. We wish
to thank A. G. Riess, E. O. Ofek, J. D. Neill, and M. Way for helpful discussions
and comments.  This work was based on data collected at the Subaru
Telescope, which is operated by the National Astronomical Observatory
of Japan, and on observations collected at the Kitt Peak National
Observatory (KPNO), National Optical Astronomical Observatories (NOAO),
which is operated by AURA, Inc., under cooperative agreement
with the National Science Foundation.
Additional data presented here were obtained at the W. M.
Keck Observatory, which is operated as a scientific partnership among
the California Institute of Technology, the University of California,
and the National Aeronautics and Space Administration; the
Observatory was made possible by the generous financial support of the
W. M. Keck Foundation.  The authors wish to recognize and acknowledge
the very significant cultural role and reverence that the summit of
Mauna Kea has always had within the indigenous Hawaiian community.  We
are most fortunate to have the opportunity to conduct observations
from this mountain.  We are grateful for support by the International
Institute for Experimental Astrophysics at Tel-Aviv University. A.V.F.'s
SN group at U.C. Berkeley is supported by US National Science Foundation
grant AST--0607485, as well as by NASA/{\it HST} grant GO--10493 from the
Space Telescope Science Institute, which is operated by AURA, Inc., 
under NASA contract NAS 5--26555. This research was supported in part 
by the National Science Foundation under Grant No. PHY--0551164.


\end{document}